\documentclass{nature_arxiv}
\usepackage{cite}
\usepackage{graphicx}
\usepackage[space]{grffile}
\usepackage{latexsym}
\usepackage{textcomp}
\usepackage{booktabs,multirow}
\usepackage{amsfonts,amsmath,amssymb}

\newif\iflatexml\latexmlfalse
\DeclareGraphicsExtensions{.pdf,.PDF,.png,.PNG,.jpg,.JPG,.jpeg,.JPEG}

\usepackage[utf8]{inputenc}
\usepackage[english]{babel}

\usepackage{hyperref}      
\usepackage{deluxetable}
\usepackage{xspace} 
\usepackage{journalnames} 

\usepackage{multicol,caption} 
\def\dataset[#1]#2{#2}%

\def\LCDM{$\Lambda$CDM\xspace}

\def\arcsec{\ensuremath{^{\prime\prime}}\xspace} 

\def\farcs{\mbox{\ensuremath{.\!\!^{\prime\prime}}}}

\def\Msun{\mbox{M$_{\odot}$}\xspace}
\def\Rsun{\mbox{R$_{\odot}$}\xspace}

\def\Ha{\mbox{H$\alpha$}\xspace}

\newcommand\forbidden[2]{[#1{\scshape{#2}}]}

\def\etacar{\ensuremath{\eta~\mbox{Car}}\xspace}
\def\etaCar{\ensuremath{\eta~\mbox{Car}}\xspace}
\def\m31n{M31N\,2008-12a\xspace}
\def\M31N{M31N\,2008-12a\xspace}

\def\HST{{\it HST}\xspace}

\def\Chandra{{\it Chandra}\xspace}

\def\Swift{{\it Swift}\xspace}

\def\spock{HFF14Spo\xspace}
\def\spockone{HFF14Spo-NW\xspace}
\def\spocktwo{HFF14Spo-SE\xspace}

\def\macs0416{MACS0416\xspace}
\def\MACS0416{MACS0416\xspace}
\def\fullmacs0416{MACS\,J0416.1-2403\xspace}
\def\Lpk{\ensuremath{L_{\rm pk}}\xspace}
\def\t2{\ensuremath{t_{2}}\xspace}

\newcommand\affiliation[1]{%
 \move@AU\move@AF%
 \begingroup%
  \@affiliation{\hspace*{2mm}#1}%
}%

\def\affil@mark#1{\textsuperscript{#1}}
\def\affile@mark@pad{0.2em}
\def\altaffilmark#1{\affil@mark{#1}}

\newcounter{affilct}
\makeatletter
\newcommand{\affilref}[1]{%
  \@ifundefined{c@#1}%
    {\newcounter{#1}%
     \setcounter{#1}{\theaffilct}%
     \refstepcounter{affilct}%
     \label{#1}%
     }{}%
  \ref{#1}%
 }
\makeatother

\makeatletter
\newcommand*\affilreftxt[2]{%
  \@ifundefined{c@#1txt}
    {\newcounter{#1txt}%
     \setcounter{#1txt}{1}
     \altaffiltext{\ref{#1}}{#2}
     }{
     }
  }
\makeatother

\newcommand{\Packard}{Packard Fellow}
\newcommand{\CalTech}{Cahill Center for Astronomy and Astrophysics, California Institute of Technology, MC 249-17, Pasadena, CA 91125, USA}
\newcommand{\BenGurion}{Ben-Gurion University of the Negev P.O.B. 653 Beer-Sheva 8410501, Israel}
\newcommand{\Cantabria}{IFCA, Instituto de F\'isica de Cantabria (UC-CSIC), Av. de Los Castros s/n, 39005 Santander, Spain}
\newcommand{\IFCA}{\Cantabria}
\newcommand{\JHU}{Department of Physics and Astronomy, The Johns Hopkins University, 3400 N. Charles St., Baltimore, MD 21218, USA}
\newcommand{\Michigan}{Department of Astronomy, University of Michigan, 1085 S. University Avenue, Ann Arbor, MI 48109, USA}
\newcommand{\UCSC}{Department of Astronomy and Astrophysics, University of California, Santa Cruz, CA 95064, USA}
\newcommand{\UCDavis}{University of California Davis, 1 Shields Avenue, Davis, CA 95616}
\newcommand{\UCLA}{Department of Physics and Astronomy, University of California, Los Angeles, CA 90095}
\newcommand{\USC}{Department of Physics and Astronomy, University of South Carolina, 712 Main St., Columbia, SC 29208, USA}
\newcommand{\TokyoRCEU}{Research Center for the Early Universe, University of Tokyo, 7-3-1 Hongo, Bunkyo-ku, Tokyo 113-0033, Japan}
\newcommand{\TokyoPhys}{Department of Physics, University of Tokyo, 7-3-1 Hongo, Bunkyo-ku, Tokyo 113-0033, Japan}
\newcommand{\TokyoIPMU}{Kavli Institute for the Physics and Mathematics of the Universe (Kavli IPMU, WPI), University of Tokyo, 5-1-5 Kashiwanoha, Kashiwa, Chiba 277-8583, Japan}
\newcommand{\TokyoAstro}{Department of Astronomy, Graduate School of Science, The University of Tokyo, 7-3-1 Hongo, Bunkyo-ku, Tokyo 113-0033, Japan}
\newcommand{\DARK}{Dark Cosmology Centre, Niels Bohr Institute, University of Copenhagen, Juliane Maries Vej 30, DK-2100 Copenhagen, Denmark} 
\newcommand{\Milan}{Dipartimento di Fisica, Universit\`a  degli Studi di Milano, via Celoria 16, I-20133 Milano, Italy}

\newcommand{\EHU}{Fisika Teorikoa, Zientzia eta Teknologia Fakultatea, Euskal Herriko Unibertsitatea UPV/EHU}
\newcommand{\Basque}{IKERBASQUE, Basque Foundation for Science, Alameda Urquijo, 36-5 48008 Bilbao, Spain}
\newcommand{\Berkeley}{Department of Astronomy, University of California, Berkeley, CA 94720-3411, USA}
\newcommand{\STScI}{Space Telescope Science Institute, 3700 San Martin Dr., Baltimore, MD 21218, USA}
\newcommand{\Ferrara}{Dipartimento di Fisica e Scienze della Terra, Universit\`{a} degli Studi di Ferrara, via Saragat 1, I-44122, Ferrara, Italy}

\newcommand{\UCSB}{Department of Physics, University of California, Santa Barbara, CA 93106-9530, USA}

\newcommand{\ASIAA}{Institute of Astronomy and Astrophysics, Academia Sinica, P.O. Box 23-141, Taipei 10617, Taiwan}

\newcommand{\Munich}{University Observatory Munich, Scheinerstrasse 1, D-81679 Munich, Germany} 

\newcommand{\Andalucia}{Instituto de Astrof\'isica de Andaluc\'ia (CSIC), E-18080 Granada, Spain}
\newcommand{\SaoPaulo}{Instituto de Astronomia, Geof\'isica e Ci\^encias Atmosf\'ericas, Universidade de S\~ao Paulo, Cidade Universit\'aria, 05508-090, S\~ao Paulo, Brazil}
\newcommand{\AMNH}{Department of Astrophysics, American Museum of Natural History, Central Park West and 79th Street, New York, NY 10024, USA}
\newcommand{\NYU}{Center for Cosmology and Particle Physics, New York University, New York, NY 10003, USA}
\newcommand{\Arizona}{Department of Astronomy, University of Arizona, Tucson, AZ 85721, USA}
\newcommand{\Rutgers}{Department of Physics and Astronomy, Rutgers, The State University of New Jersey, Piscataway, NJ 08854, USA}

\newcommand{\LCOGT}{Las Cumbres Observatory Global Telescope Network, 6740 Cortona Dr., Suite 102, Goleta, California 93117, USA}

\newcommand{\AIP}{Leibniz-Institut f\"ur Astrophysik Potsdam (AIP), An der Sternwarte 16, 14482 Potsdam, Germany}
\newcommand{\CEA}{Centre for Extragalactic Astronomy, Department of Physics, Durham University, Durham DH1 3LE, U.K.}
\newcommand{\ICC}{Institute for Computational Cosmology, Durham University, South Road, Durham DH1 3LE, U.K.}
\newcommand{\ACRU}{Astrophysics and Cosmology Research Unit, School of Mathematical Sciences, University of KwaZulu-Natal, Durban 4041, South Africa}
\newcommand{\MPIA}{Max-Planck-Institut f{\"u}r Astrophysik, Karl-Schwarzschild-Str.~1, 85748 Garching, Germany}
\newcommand{\Garching}{Physik-Department, Technische Universit\"at M\"unchen, James-Franck-Stra\ss{}e~1, 85748 Garching, Germany}
\newcommand{\Riverside}{Department of Physics and Astronomy, University of California, Riverside, CA 92521, USA}
\newcommand{\UCRiverside}{\Riverside}
\newcommand{\CfA}{Harvard/Smithsonian Center for Astrophysics, Cambridge, MA 02138, USA}
\newcommand{\Minnesota}{School of Physics and Astronomy, University of Minnesota, 116 Church Street SE, Minneapolis, MN 55455, USA}
\newcommand{\Lyon}{Universit\'e Lyon, Univ Lyon1, Ens de Lyon, CNRS,
  Centre de Recherche Astrophysique de Lyon UMR5574, F-69230,
  Saint-Genis-Laval, France}

\title{Two Peculiar Fast Transients in a Strongly Lensed Host Galaxy}

\author{
S.~A.~Rodney\altaffilmark{\affilref{USC}},
I.~Balestra\altaffilmark{\affilref{Munich}},
M.~Brada\v{c}\altaffilmark{\affilref{UCDavis}},
G.~Brammer\altaffilmark{\affilref{STScI}},
T.~Broadhurst\altaffilmark{\affilref{EHU},\affilref{Basque}},
G.~B.~Caminha\altaffilmark{\affilref{Ferrara}},
G.~Chiriv{\`i}\altaffilmark{\affilref{MPIA}},
J.~M.~Diego\altaffilmark{\affilref{IFCA}},
A.~V.~Filippenko\altaffilmark{\affilref{Berkeley}},
R.~J.~Foley\altaffilmark{\affilref{UCSC}},
O.~Graur\altaffilmark{\affilref{NYU},\affilref{AMNH},\affilref{CfA}},
C.~Grillo\altaffilmark{\affilref{Milan},\affilref{DARK}},
S.~Hemmati\altaffilmark{\affilref{CalTech}},
J.~Hjorth\altaffilmark{\affilref{DARK}},
A.~Hoag\altaffilmark{\affilref{UCDavis}},
M.~Jauzac\altaffilmark{\affilref{CEA},\affilref{ICC},\affilref{ACRU}},
S.~W.~Jha\altaffilmark{\affilref{Rutgers}},
R.~Kawamata\altaffilmark{\affilref{TokyoAstro}},
P.~L.~Kelly\altaffilmark{\affilref{Berkeley}},
C.~McCully\altaffilmark{\affilref{LCOGT},\affilref{UCSB}},
B.~Mobasher\altaffilmark{\affilref{UCRiverside}},
A.~Molino\altaffilmark{\affilref{SaoPaulo},\affilref{Andalucia}},
M.~Oguri\altaffilmark{\affilref{TokyoRCEU},\affilref{TokyoPhys},\affilref{TokyoIPMU}},
J.~Richard\altaffilmark{\affilref{Lyon}},
A.~G.~Riess\altaffilmark{\affilref{JHU},\affilref{STScI}},
P.~Rosati\altaffilmark{\affilref{Ferrara}},
K.~B.~Schmidt\altaffilmark{\affilref{UCSB},\affilref{AIP}},
J.~Selsing\altaffilmark{\affilref{DARK}},
K.~Sharon\altaffilmark{\affilref{Michigan}},
L.-G.~Strolger\altaffilmark{\affilref{STScI}},
S.~H.~Suyu\altaffilmark{\affilref{MPIA},\affilref{ASIAA},\affilref{Garching}},
T.~Treu\altaffilmark{\affilref{UCLA},\affilref{Packard}},
B.~J.~Weiner\altaffilmark{\affilref{Arizona}},
L.~L.~R.~Williams\altaffilmark{\affilref{Minnesota}} \&
A.~Zitrin\altaffilmark{\affilref{BenGurion}}
}

\def\makeaffil{
\begin{affiliations}
 \item \USC
 \item \Munich
 \item \UCDavis
 \item \STScI
 \item \EHU
 \item \Basque
 \item \Ferrara
 \item \MPIA
 \item \IFCA
 \item \Berkeley
 \item \UCSC
 \item \NYU
 \item \AMNH
 \item \CfA
 \item \Milan
 \item \DARK
 \item \CalTech
 \item \CEA
 \item \ICC
 \item \ACRU
 \item \Rutgers
 \item \TokyoAstro
 \item \LCOGT
 \item \UCSB
 \item \UCRiverside
 \item \SaoPaulo
 \item \Andalucia
 \item \TokyoRCEU
 \item \TokyoPhys
 \item \TokyoIPMU
 \item \Lyon
 \item \JHU
 \item \AIP
 \item \Michigan
 \item \ASIAA
 \item \Garching
 \item \UCLA
 \item \Packard
 \item \Arizona
 \item \Minnesota
 \item \BenGurion
\end{affiliations}
}

\begin{document}

\maketitle

\vspace{10pt}

\begin{abstract}
A massive galaxy cluster can serve as a magnifying glass for distant
stellar populations, with strong gravitational lensing exposing details
in the lensed background galaxies that would otherwise be undetectable.
The \fullmacs0416 cluster (hereafter \macs0416) is one of the most
efficient lenses in the sky, and in 2014 it was observed with high-cadence
imaging from the Hubble Space Telescope (\HST). Here we describe two
unusual transient events that appeared behind \macs0416 in a strongly lensed galaxy at redshift
$z=1.0054\pm0.0002$. These transients---designated
\spockone and \spocktwo and collectively nicknamed ``Spock''---were
faster and fainter than any supernova (SN), but significantly more luminous
than a classical nova. They reached peak luminosities of $\sim10^{41}$
erg s$^{-1}$ ($M_{AB}<-14$ mag) in $\lesssim$5 rest-frame days, then faded
below detectability in roughly the same time span.  Models of the
cluster lens suggest that these events may be {\it spatially}
coincident at the source plane, but are most likely not {\it
  temporally} coincident.  We find that \spock can be explained as a
luminous blue variable (LBV), a recurrent nova (RN), or a pair of stellar
microlensing events.  To distinguish between these hypotheses will
require a clarification of the positions of nearby critical curves,
along with high-cadence monitoring of the field that could detect new
transient episodes in the host galaxy.
\end{abstract}

\vspace{10pt}

\makeaffil


 When a star explodes or a relativistic jet erupts from near the edge
of a black hole, the event can be visible across many billions of
light-years.  Such extremely luminous astrophysical transients as
supernovae (SNe), gamma-ray bursts, and quasars are powerful tools for
probing cosmic history and sampling the matter and energy content of
the universe.  Less energetic transients generated by the tumultuous
atmospheres of massive stars or the interactions of close stellar
binaries are also very valuable for understanding stellar evolution
and the physical processes that lead to stellar explosions.  However,
the lower luminosity of such events makes them accessible only in
the local universe, and consequently our census of peculiar transients
at the stellar scale is still highly incomplete.


Although recent surveys are beginning to discover progressively more 
categories of rapidly changing optical
transients\cite{Kasliwal:2011a,Drout:2014}, most programs remain
largely insensitive to transients with peak brightness and timescales
comparable to the \spock events\cite{Berger:2013b}.  Future wide-field
observatories such as the Large Synoptic Survey
Telescope\cite{Tyson:2002} will be much more efficient at discovering
such transients, and can be expected to reveal many new categories of
astrophysical transients.

As shown in Figure~\ref{fig:SpockDetectionImages}, the \spock events
appeared in \HST imaging collected as part of
the Hubble Frontier Fields (HFF) survey\cite{Lotz:2017}, a multi-cycle
program for deep imaging of 6 massive galaxy clusters and associated
``blank sky'' fields observed in parallel.  \HST is not an efficient
wide-field survey telescope, and the HFF survey was not designed with
the discovery of peculiar extragalactic transients as a core
objective.  However, the HFF program has unintentionally opened an
effective window of discovery for such events.  Very faint sources at
relatively high redshift ($z\gtrsim1$) in these fields are made
detectable by the substantial gravitational lensing magnification from
the foreground galaxy clusters.  Very rapidly evolving sources are
also more likely to be found, owing to the necessity of a rapid cadence
for repeat imaging in the HFF program.

\begin{figure*}[tbp]
\begin{center}
\includegraphics[width=1\textwidth]{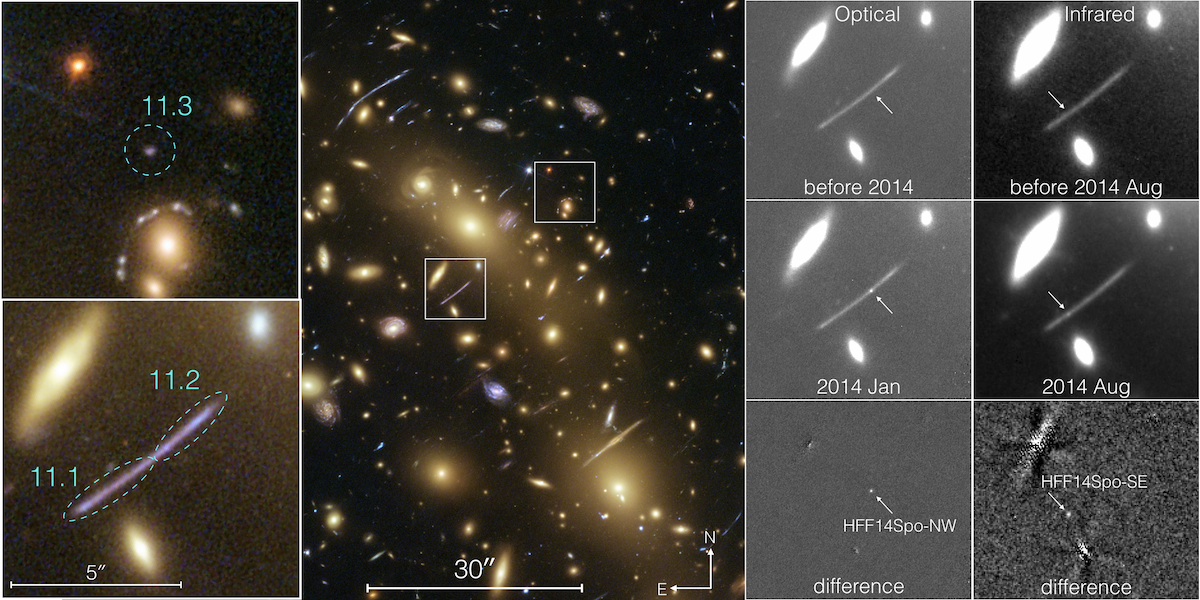}
\caption{ \protect\input{detection_image_caption.tex}}
\end{center}
\end{figure*}

\begin{deluxetable}{lccccc}
  \tablewidth{0.7\textwidth}
  \tablecolumns{6}
  \tablecaption{Lens model predictions for time delays and
    magnifications at the observed locations of the \spock
    transients. \label{tab:LensModelPredictions}}
  \tablehead{
    \colhead{Model} &
    \colhead{$\lvert\mu_{\rm NW}\rvert$} & \colhead{$\lvert\mu_{\rm SE}\rvert$} &
    \colhead{$\lvert\mu_{11.3}\rvert$} &  
    \colhead{$\Delta t_{\rm NW:SE}$} & \colhead{$\Delta t_{\rm NW:11.3}$} \\
    \colhead{} & \colhead{} & \colhead{} & \colhead{} & \colhead{(days)} & \colhead{(years)}}
\startdata
CATS &  196$^{+140}_{-53}$ & 46$^{+2}_{-1}$ & 3.3$^{+0.0}_{-0.0}$ & -1.7$^{+2.0}_{-1.9}$ & -3.7$^{+0.1}_{-0.2}$\\[0.5em]
GLAFIC & 29$^{+43}_{-10}$ &  84$^{+103}_{-38}$ &  3.0$^{+0.2}_{-0.2}$ &  4.1$^{+5.5}_{-3.4}$ &  -5.0$^{+0.5}_{-0.6}$\\[0.5em]
GLEE & 182$^{+203}_{-83}$ &  67$^{+31}_{-16}$ &  2.9$^{+0.1}_{-0.1}$ &  36$^{+6}_{-7}$ &  -6.1$^{+0.3}_{-0.2}$\\[0.5em]
GRALE & 13$^{+11}_{-6}$ & 12$^{+9}_{-5}$ & 3.1$^{+2.2}_{-0.9}$ & -10$^{+1}_{-7}$ & -2.5$^{+1.0}_{-3.1}$ \\[0.5em]
SWunited & 38$\pm8$ & 13$\pm1$ & 2.9 $\pm0.1$ & \nodata & \nodata \\[0.5em]
WSLAP$^{+}$ & 35$\pm$20 & 30$\pm$20 & \nodata & -48$\pm$10 & 0.8  \\[0.5em]
ZLTM & 103$^{+48}_{-40}$ & 32$^{+8}_{-10}$ & 3.5$\pm$0.3  & 43$^{+12}_{-10}$ & -3.7$\pm0.3$\\
\enddata
\tablecomments{ Each lens model is identified by the name of the
  modeling team or tool. Time delays give the predicted delay relative
  to an appearance in the NW host image, 11.2. Positive (negative)
  values indicate the NW image is the leading (trailing) image of the
  pair.  The observed time lag between the NW and SE events was
  $\Delta t_{\rm NW:SE}=234\pm6$ days.}
  \label{tab:LensModelPredictions}
\end{deluxetable}


\begin{figure*}[tbp]
  \begin{center}
    \includegraphics[width=\textwidth]{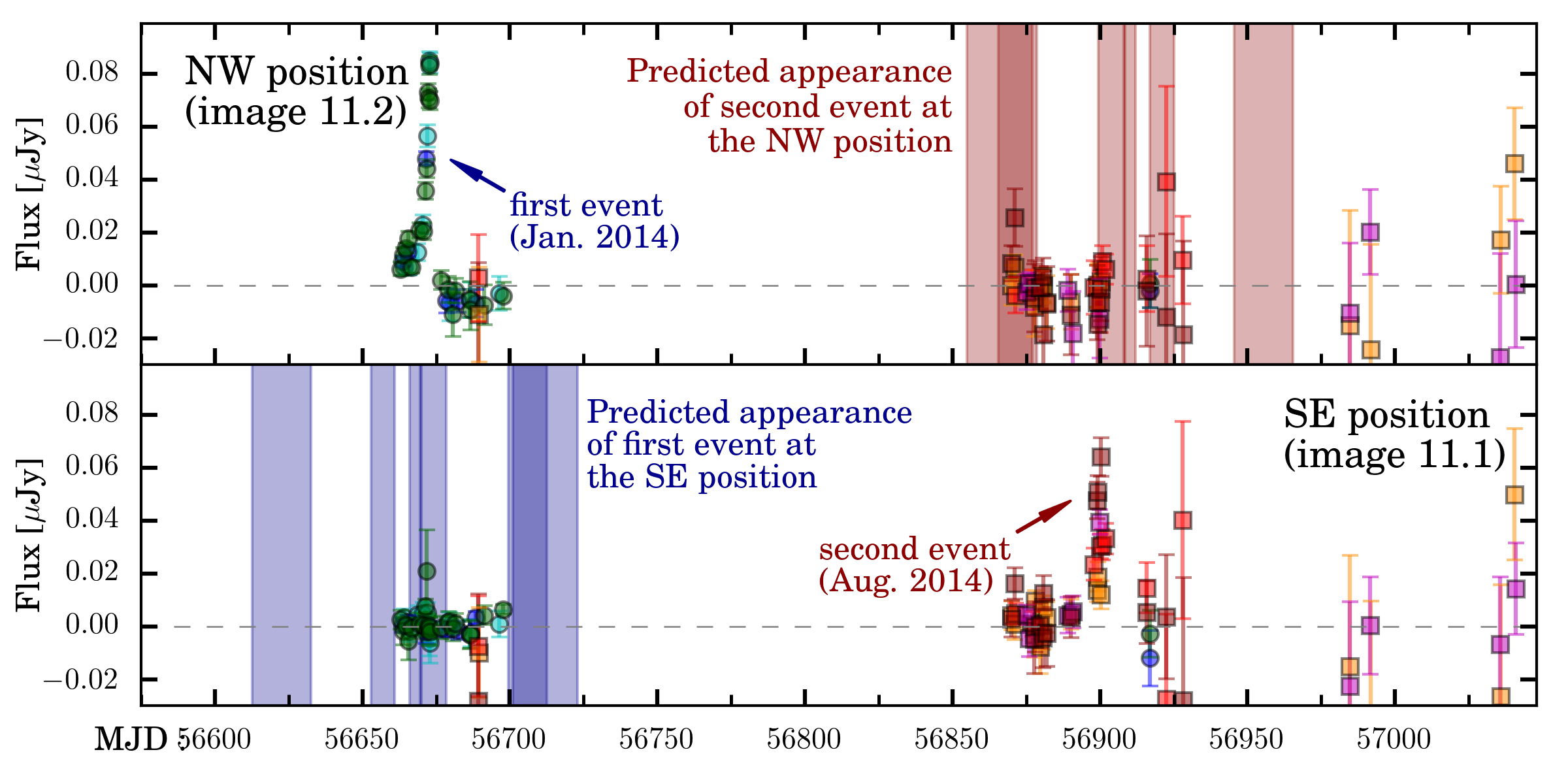}
    \caption{\protect\input{spock_predictions_caption.tex}}
  \end{center}
\end{figure*}

\begin{figure*}[tbp]
  \begin{center}
    \includegraphics[width=\textwidth]{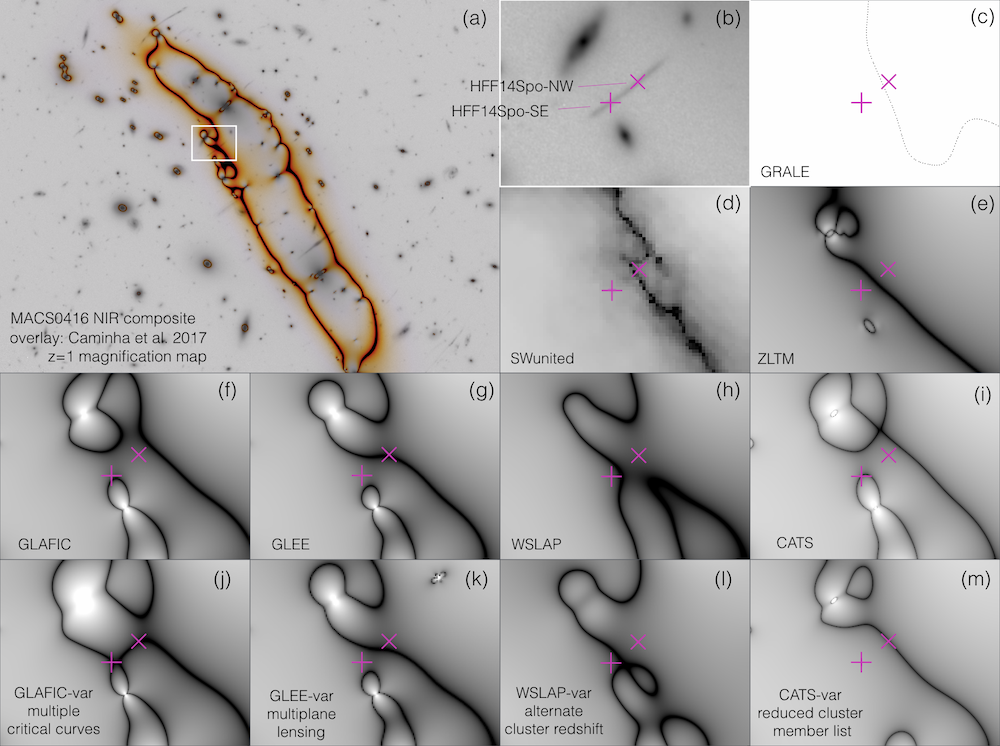}
    \caption{\protect\input{spock_critical_curves_caption.tex}}
  \end{center}
\end{figure*}

\begin{figure*}[tbp]
\begin{center}
\includegraphics[width=1\textwidth]{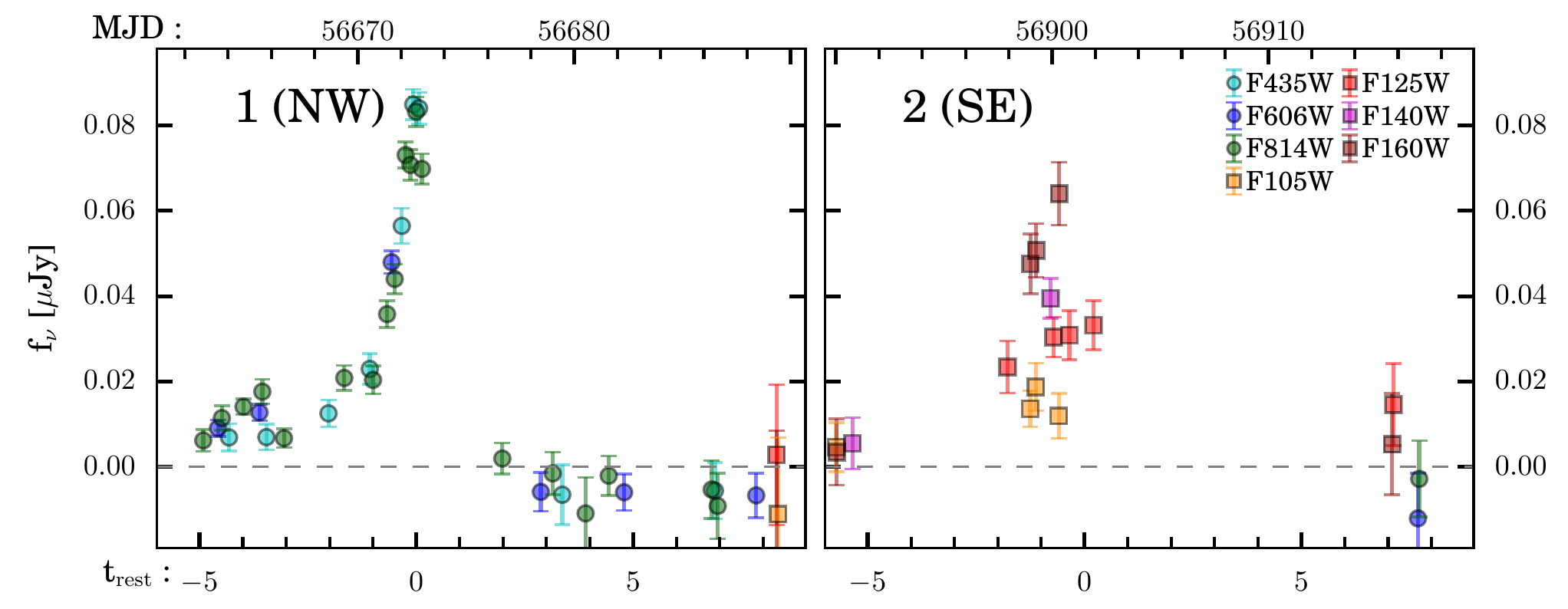}
\caption{ \protect\input{spock_lightcurves_caption.tex}}
\end{center}
\end{figure*}

\begin{figure*}[tbp]
\begin{center}
\includegraphics[width=0.48\textwidth]{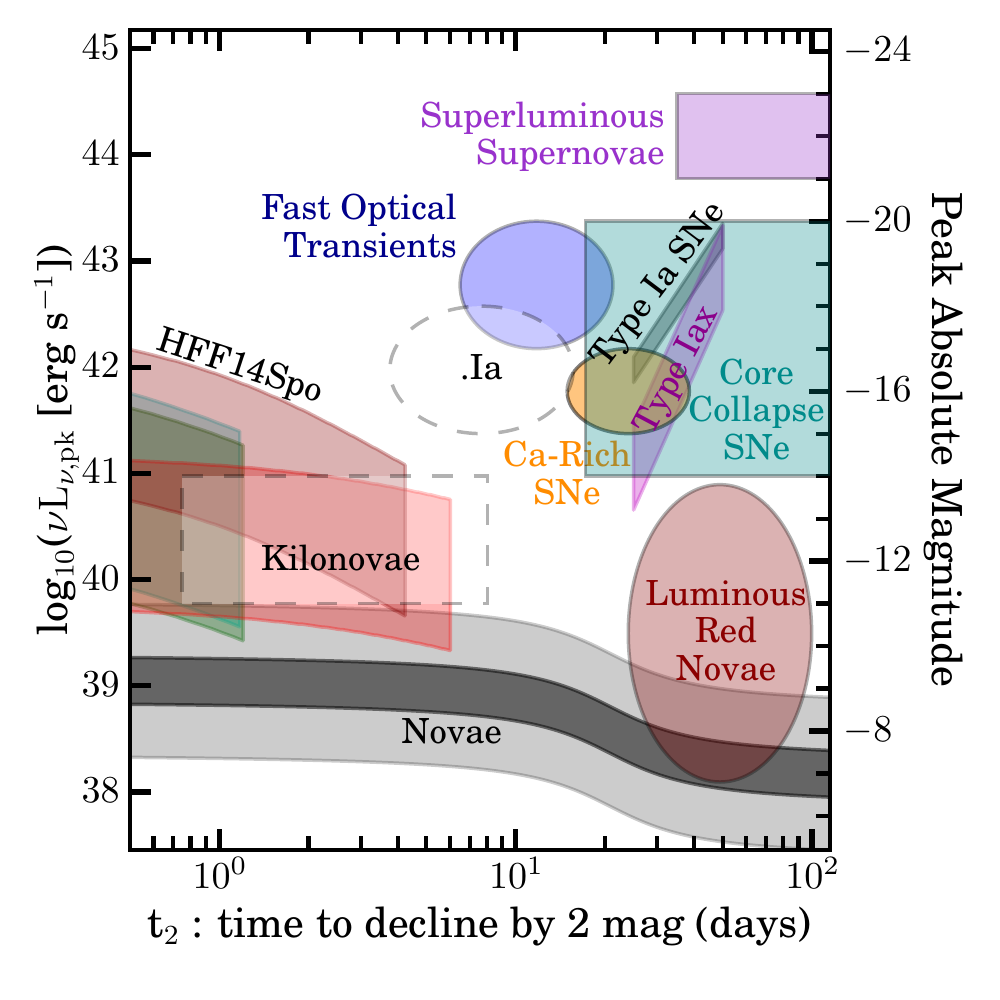}
\includegraphics[width=0.48\textwidth]{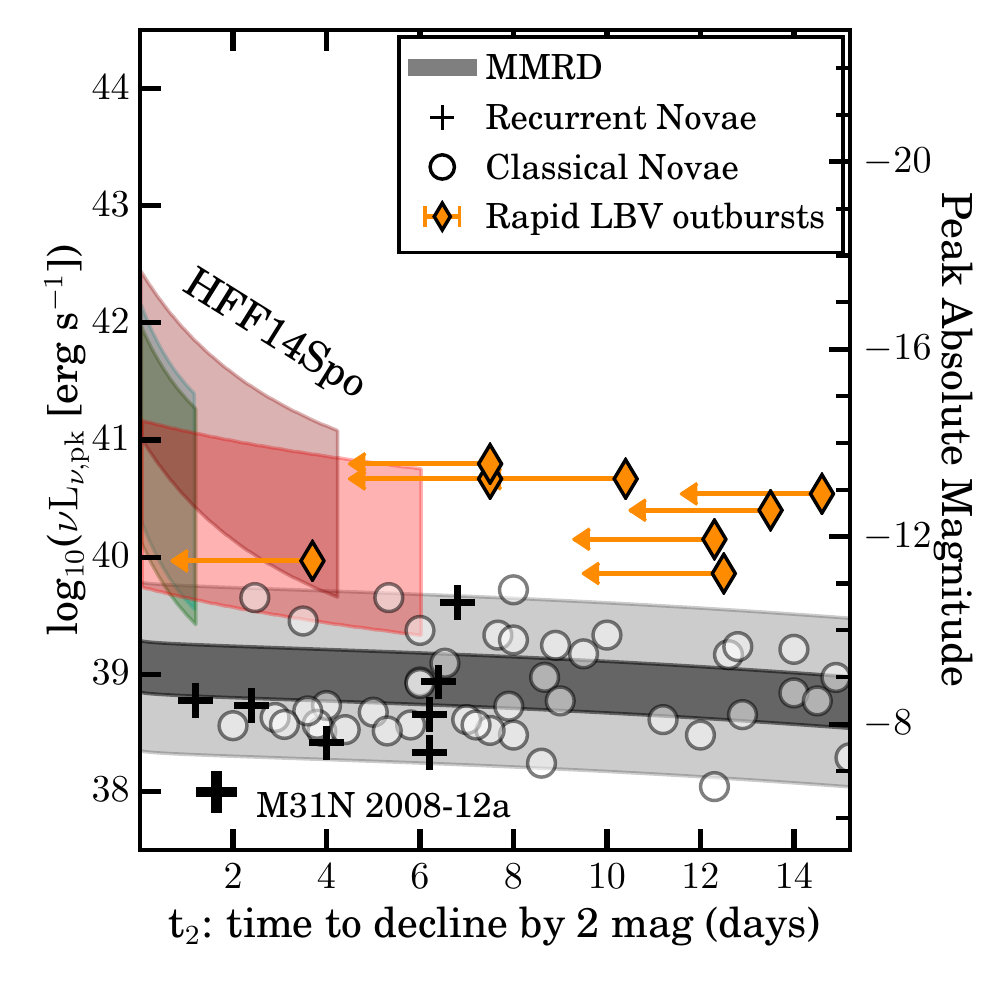}
\caption{ \protect\input{peakluminosity_vs_declinetime_caption.tex}}
\end{center}
\end{figure*}

\section{Results}\label{sec:Results}

To evaluate the impact of gravitational lensing from the \MACS0416
cluster on the observed light curves and the timing of these two
events, we use seven independently constructed cluster mass models.
These models indicate that the gravitational time delay between the
\spockone location and the \spocktwo location is $<$60 days
(Table~\ref{tab:LensModelPredictions}).  This falls far short of the
observed 223 day span between the two events, suggesting that
\spocktwo is not a time-delayed image of the \spockone event.  As
shown in Figure~\ref{fig:SpockDelayPredictions}, \spockone and
\spocktwo are inconsistent with these predicted time delays if one
assumes that they are delayed images of a single event.  However, if
these were independent events, then a time delay on the order of tens
of days between image 11.1 and 11.2 could have resulted in
time-delayed events that were missed by the \HST imaging of this
field.

The models also predict absolute magnification values between about
$\mu=10$ and $\mu=200$ for both events. This wide range is due
primarily to the close proximity of the lensing critical curve (the
region of theoretically infinite magnification) for sources at $z=1$.
The lensing configuration consistently adopted for this cluster
assumes that the arc comprises two mirror images of the host galaxy
(labeled 11.1 and 11.2 in
Figure~\ref{fig:SpockDetectionImages})\cite{Zitrin:2013a, Jauzac:2014,
  Johnson:2014, Richard:2014, Diego:2015a, Grillo:2015, Hoag:2016,
  Sebesta:2016, Caminha:2017}.  This implies that a single critical
curve passes roughly midway between the two \spock locations.  The
location of the critical curve varies significantly among the models
(Figure~\ref{fig:SpockCriticalCurves}), and is sensitive to many
parameters that are poorly constrained. We find that it is possible to
make reasonable adjustments to the lens model parameters so that the
critical curve does not bisect the \spock host arc, but instead
intersects both of the \spock locations (see Supplementary
Note~\ref{sec:LensModelVariations}).  Such lensing configurations can
qualitatively reproduce the observed morphology of the \spock host
galaxy, but they are disfavored by a purely quantitative assessment of
the positional strong-lensing constraints.

\subsection{Ruling Out Common Astrophysical Transients.}

There are several categories of astrophysical transients that can be
rejected based solely on characteristics of the \spockone and
\spocktwo light curves, shown in Figure~\ref{fig:LightCurves}. Neither
of the \spock events is {\it periodic}, as expected for stellar
pulsations such as Cepheids, RR Lyrae, or Mira variables. Stellar
flares can produce rapid optical transient phenomena, but the total
energy released by even the most extreme stellar
flare\cite{Karoff:2016} falls far short of the observed energy release
from the \spock transients. We can also rule out active galactic
nuclei (AGN), which are disfavored by the quiescence of the \spock
sources between the two observed episodes and the absence of any of
the broad emission lines that are often observed in AGN.
Additionally, no x-ray emitting point source was detected in 7 epochs
from 2009 to 2014, including \Chandra X-ray Space Telescope imaging
that was coeval with the peak of infrared emission from \spocktwo.

Dynamically induced stellar collisions or close interactions in a
dense stellar cluster\citep{Fregeau:2004} could in principle produce a
series of optical transients. Similarly, the collision of a jovian
planet with a main sequence star\cite{Metzger:2012,Yamazaki:2017} or a
terrestrial planet with a white dwarf star\cite{Di-Stefano:2015} could
generate an optical transient with a peak luminosity comparable to
that observed for the \spock events, although it is unclear whether
the UV/optical emission could match the observed \spock light curves.
These scenarios warrant further scrutiny, so that predictions of the
light curve shape and anticipated rates can be more rigorously
compared to the \spock observations.

Many types of stellar explosions can generate isolated transient
events, and a useful starting point for classification of such objects
is to examine their position in the phase space of peak luminosity
(\Lpk) versus decline time\cite{Kulkarni:2007}.
Figure~\ref{fig:PeakLuminosityDeclineTime} shows our two-dimensional
constraints on \Lpk and the decline timescale \t2 (the time over which
the transient declines by 2 mag) for the \spock events,
accounting for the range of lensing magnifications ($10<\mu<200$)
derived from the cluster lens models.  The \spockone and \spocktwo
events are largely consistent with each other, and if both events are
representative of a single system (or a homogeneous class) then the
most likely peak luminosity and decline time (the region with the most
overlap) would be $L_{\rm pk}\approx10^{41}$ erg s$^{-1}$ and $t_2\approx1$
day.

The relatively low peak luminosities and the very rapid rise and fall
of both \spock light curves are incompatible with all categories of
stellar explosions for which a significant sample of observed events
exists.  This includes the common Type Ia SNe and core-collapse SNe,
as well as the less well-understood classes of superluminous
SNe\cite{Gal-Yam:2012}, Type Iax SNe\citep{Foley:2013a}, fast optical
transients\cite{Drout:2014}, Ca-rich SNe\cite{Kasliwal:2012}, and
luminous red novae\cite{Kulkarni:2007}.  

The SN-like transients that come closest to matching the observed
light curves of the two \spock events are the ``kilonova'' class and
the ``.Ia'' class.  Kilonovae are a category of optical/near-infrared
transients that may be generated by the merger of a neutron star (NS)
binary\cite{Li:1998, Tanvir:2013, Jin:2016}.  The .Ia class is
produced by He shell explosions that are expected to arise from AM
Canum Venaticorum (AM CVn) binary star systems undergoing He mass
transfer onto a white dwarf primary star\cite{Bildsten:2007}.  The
\spock light curves exhibited a slower rise time than is expected for
a kilonova event\cite{Barnes:2013, Kasen:2015}, and a faster decline
time than is anticipated for a .Ia event\cite{Shen:2010}.

Another problem for all of these catastrophic stellar explosion models
is that they cannot explain the appearance of {\it repeated} transient
events.  The kilonova progenitor systems are completely disrupted at
explosion, as is the case for all normal SN explosions.  For .Ia
events, even if an AM CVn system could produce repeated He shell
flashes of similar luminosity, the period of recurrence would be
$\sim10^5$ yr, making these effectively non-recurrent sources.  Models
invoking a stellar merger or the collision of a planet with its parent
star have a similar difficulty.  In these cases the star may survive
the encounter, but the rarity of these collision events makes it
highly unlikely to detect two such transients from the same galaxy in
a single year.

Although the two events were most likely not {\it temporally}
coincident, all of our lens models indicate that it is entirely
plausible for the two \spock events to be {\it spatially} coincident:
a single location at the source plane can be mapped to both \spock
locations to within the positional accuracy of the model
reconstructions ($\sim$0.6\arcsec in the lens plane). This is
supported by the fact that the host-galaxy colors and spectral indices
at each \spock location are indistinguishable within the uncertainties
(see Supplementary Figure~\ref{fig:HostProperties} and Supplementary
Table~\ref{tab:HostProperties}).  Thus, to accommodate all of the
observations of the \spock events with a single astrophysical source,
we turn to two categories of stellar explosion that are sporadically
recurrent: luminous blue variables (LBVs) and recurrent novae (RNe).

\subsection{Luminous Blue Variable.}

The transient sources categorized as LBVs are the result of eruptions
or explosive episodes from massive stars ($>10$ \Msun).
The class is exemplified by examples such as P Cygni, $\eta$ Carinae
(\etaCar), and S Doradus\cite{Smith:2011b, Kochanek:2012}.  Although
most giant LBV eruptions have been observed to last much longer than
the \spock events\cite{Smith:2011b}, some LBVs have exhibited repeated
rapid outbursts that are broadly consistent with the very fast \spock
light curves (see Supplementary
Figure~\ref{fig:LBVLightCurveComparison}). Because of this common
stochastic variability, the LBV hypothesis does not have any trouble
accounting for the \spock events as two separate episodes.

Two well-studied LBVs that provide a plausible match to the observed
\spock events are ``SN 2009ip''\cite{Maza:2009} and
NGC3432-LBV1\cite{Pastorello:2010}.  Both exhibited multiple brief
transient episodes over a span of months to years.  Unfortunately,
for these outbursts we have only upper limits on the decline
timescale, $t_2$, owing to the relatively sparse photometric sampling.
Recent studies have shown that SN 2009ip-like LBV transients have remarkably
similar light curves, leading up to a final terminal SN
explosion\cite{Kilpatrick:2017, Pastorello:2017}.
Figure~\ref{fig:PeakLuminosityDeclineTime}b shows that both \spock
events are consistent with the observed luminosities and decline times
of these fast and bright LBV outbursts -- though the \spock events
would be among the most rapid and most luminous LBV eruptions ever
seen.

In addition to those relatively short and very bright giant eruptions,
most LBVs also commonly exhibit a slower underlying variability. P
Cygni and \etaCar, for example, slowly rose and fell in brightness by
$\sim$1 to 2 mag over a timespan of several years before and after
their historic giant eruptions.  Such variation has not been detected
at the \spock locations. Nevertheless, given the broad
range of light-curve behaviors seen in LBV events, we cannot reject
this class as a possible explanation for the \spock system.

The total radiated energy of the \spock events is in the range
$10^{44}<E_{\rm rad}<10^{47}$ erg (see Methods), which falls well
within the range of plausible values for a major LBV outburst.  From
this measurement we can derive constraints on the luminosity of the
progenitor star, by assuming that the energy released is generated
slowly in the stellar interior and is in some way ``bottled up'' by
the stellar envelope, before being released in a rapid mass ejection
(see Methods).  With this approach we a quiescent luminosity of
$L_{\rm qui}\approx10^{39.5}$~erg~s$^{-1}$ ($M_V\approx-10$ mag).  This value is
fully consistent with the expected range for LBV progenitor stars
(e.g., \etacar has $M_V\approx-12$ mag and the faintest known LBV progenitors
such as SN 2010dn have $M_V\approx-6$ mag).

\subsection{Recurrent Nova.}\label{sec:RNe}

Novae occur in binary systems in which a white dwarf star accretes
matter from a less massive companion, leading to a burst of nuclear
fusion in the accreted surface layer that causes the white dwarf to
brighten by several orders of magnitude, but does not completely
disrupt the star. The mass transfer from the companion to the white
dwarf may restart after the explosion, so the cycle may begin again
and repeat after a period of months or years.  When this recurrence
cycle is directly observed, the object is classified as a recurrent
nova (RN).

The light curves of many RN systems in the Milky Way are similar in
shape to the \spock episodes, exhibiting a sharp rise ($<10$ days in
the rest-frame) and a similarly rapid decline (see Supplementary
Information and Supplementary
Figure~\ref{fig:RecurrentNovaLightCurveComparison}).  This is
reflected in Figure~\ref{fig:PeakLuminosityDeclineTime}, where novae
are represented by a grey band that traces the empirical constraints
on the maximum magnitude vs.\ rate of decline (MMRD) relation for
classical novae\cite{DellaValle:1995, Downes:2000}.

The RN model can provide a natural explanation for having two separate
explosions that are coincident in space but not in time.  However, the
recurrence timescale for \spock in the rest frame is $120\pm30$ days,
which would be a singularly rapid recurrence period for a RN system.
The RNe in our own Galaxy have recurrence timescales of 10--98
years\cite{Schaefer:2010}.  The fastest measured recurrence timescale
belongs to M31N 2008-12a, which has exhibited a new outburst every
year from 2008 through 2016\cite{Tang:2014, Darnley:2016}
Although this M31 record-holder demonstrates that very rapid
recurrence is possible, classifying \spock as a RN would still require
a very extreme mass-transfer rate to accommodate the $<1$ year
recurrence.

Another major concern with the RN hypothesis is that the two \spock
events are substantially brighter than all known novae---perhaps by as
much as 2 orders of magnitude.  This is exacerbated by the
observational and theoretical evidence indicating that
rapid-recurrence novae have less energetic eruptions\cite{Yaron:2005}
(see Supplementary Information and Supplementary Figure
\ref{fig:RecurrentNovaRecurrenceComparison}).
Although the RN model
is not strictly ruled out, we can deduce that if the \spock transients
are caused by a single RN system, then that progenitor system would be
among the most extreme white dwarf binary systems yet known.


\subsection{Microlensing.}\label{sec:MicroLensing}

In the presence of strong gravitational lensing it is possible to
generate a transient event from lensing effects alone.  In this case
the background source has a steady luminosity but the relative motion
of the source, lens, and observer causes the magnification of that
source (and therefore the apparent brightness) to change rapidly with
time.  An isolated strong lensing event with a rapid timescale can be
generated when a background star crosses over a lensing caustic (the
mapping of the critical curve back on to the source plane).  In the
case of a star crossing the caustic of a smooth lensing potential, the
amplification of the source flux would increase (decrease) with a
characteristic $t^{-1/2}$ profile as it moves toward (away from) the
caustic. This slowly evolving light curve then transitions to a very
sharp decline (rise) when the star has moved to the other side of the
caustic\cite{Schneider:1986, MiraldaEscude:1991}.  With a more complex
lens comprising many compact objects, the light curve would exhibit a
superposition of many such sharp peaks\cite{Lewis:1993, Diego:2017}.

The peculiar transient MACS J1149 LS1, observed behind the Hubble
Frontier Fields cluster MACS J1149.6+2223, has been proposed as the
first observed example of such a stellar caustic crossing
event\cite{Kelly:2017}. Such events may be expected to appear more
frequently in strongly lensed galaxies that have small angular
separation from the center of a massive cluster. In such a situation,
our line of sight to the lensed background galaxy passes through a
dense web of overlapping microlenses caused by the intracluster stars
distributed around the center of the cluster. This has the effect of
``blurring'' the magnification profile across the cluster critical
curve, making it more likely that a single (and rare) massive star in
the background galaxy gets magnified by the required factor of
$\sim10^5$ to become visible as a transient caustic-crossing event.
On this basis the \spock host-galaxy images are suitably positioned
for caustic-crossing transients, as they are seen through a relatively
high density of intracluster stars (see Methods)---comparable to that
observed for the MACS J1149 LS1 transient.

The characteristic timescale of a canonical caustic-crossing event
would be on the order of hours or days (see Supplementary
Information), which is comparable to the timescales observed for the
\spock events. Gravitational lensing is achromatic as long as the size
of the source is consistent across the spectral energy distribution
(SED).  This means that the color of a caustic-crossing transient will
be roughly constant.  Using simplistic linear interpolations of the
observed light curves (see Methods), we find that the inferred color
curves for both \spock events are marginally consistent with this
expectation of an unchanging color (Supplementary
Figure~\ref{fig:ColorCurves}).

In the baseline lensing configuration adopted above---where a single
critical curve subtends the \spock host galaxy arc---these events
cannot plausibly be explained as stellar caustic crossings, because
neither transient is close enough to the single critical curve to
reach the required magnifications of $\mu\approx10^6$.  Some of our lens
models can, however, be modified so that instead of just two host
images, the lensed galaxy arc is made up of many more images of the
host, with multiple critical curves subtending the arc where the
\spock events appeared (Figure~\ref{fig:SpockCriticalCurves}).
If this alternative lensing
situation is correct, then similar microlensing transients would be
expected to appear at different locations along the host-galaxy arc,
instigated by new caustic-crossing episodes from different stars in
the host galaxy.

\subsection{The Rate of Similar Transients.}\label{sec:Rates}

Although we lack a definitive classification for these events, we can
derive a simplistic estimate of the rate of \spock-like transients by
counting the number of strongly lensed galaxies in the HFF clusters
that have sufficiently high magnification that a source with
$M_{V}=-14$ mag would be detected in \HST imaging. There are only six
galaxies that satisfy that criteria, all with $0.5<z<1.5$
(Methods).  Each galaxy was observed by the high-cadence HFF program
for an average of 80 days.  Treating \spockone and \spocktwo as
separate events leads to a very rough rate estimate of 1.5 \spock-like
events per galaxy per year.

Derivation of a volumetric rate for such events would require a
detailed analysis of the lensed volume as a function of redshift, and
is beyond the scope of this work. Nevertheless, a comparison to rates
of similar transients in the local universe can inform our assessment
of the likelihood that the \spock events are unrelated.  A study of
very fast optical transients with the Pan-STARRS1 survey derived a
rate limit of $\lesssim0.05$ Mpc$^{-3}$ yr$^{-1}$ for transients
reaching $M\approx -14$ mag on a timescale of $\sim$1
day\citet{Berger:2013b}.  This limit, though several orders of
magnitude higher than the constraints on novae or SNe, is sufficient
to make it exceedingly unlikely that two unrelated fast optical
transients would appear in the same galaxy in a single year.
Furthermore, we have observed no other transient events with similar
luminosities and light curve shapes in high-cadence surveys of five
other Frontier Fields clusters. Indeed, all other transients detected
in the primary HFF survey have been fully consistent with normal SNe.
Thus, we have no evidence to suggest that transients of this kind are
common enough to be observed twice in a single galaxy in a single
year.

\section{Discussion}\label{sec:Discussion}

We have examined three plausible explanations for the \spock events:
(1) they were separate rapid outbursts of an LBV star, (2) they were
surface explosions from a single RN, or (3) they were each caused by
the rapidly changing magnification as two unrelated massive stars
crossed over lensing caustics. We cannot make a definitive choice
between these hypotheses, principally due to the scarcity of
observational data and the uncertainty in the location of the
lensing critical curves.

If there is just a single critical curve for a source at $z=1$ passing
between the two \spock locations, then our preferred explanation for
the \spock events is that we have observed two distinct eruptive
episodes from a massive LBV star.
In this scenario, the \spock LBV system would most likely
have exhibited multiple eruptions over the last few years, but most of
them were missed, as they landed within the large gaps of the \HST
Frontier Fields imaging program.
The \spock events would be extreme LBV outbursts in several
dimensions, and should add a useful benchmark for the outstanding
theoretical challenge of developing a comprehensive physical model
that accommodates both the \etacar-like great eruptions and the S
Dor-type variation of LBVs.

If instead the \macs0416 lens has multiple critical curves that
intersect both \spock locations, then the third proposal of a
microlensing-generated transient would be preferred.  Stellar caustic
crossings have not been observed before, but the analysis of a likely
candidate behind the MACSJ1149 cluster\citep{Kelly:2017} suggests that massive cluster
lenses may generate such events more frequently than previously
expected\citep{Kelly:2017, Diego:2017}. To resolve the uncertainty of
the \spock classification will require refinement of the lens models
to more fully address systematic biases and more tightly constrain the
path of the critical curve.  High-cadence monitoring of the \macs0416
field would also be valuable, as it could catch future LBV eruptions
or microlensing transients at or near these locations.


\clearpage
\begin{methods}
 
 \subsection{Discovery.}\label{sec:Discovery}

The transient \spock\ was discovered in \HST imaging collected as part
of the Hubble Frontier Fields (HFF) survey (HST-PID: 13496, PI: Lotz),
a multi-cycle program observing 6 massive galaxy clusters and
associated ``blank sky'' parallel fields\cite{Lotz:2017}.  Several
\HST observing programs have provided additional observations
supplementing the core HFF program.  One of these is the FrontierSN
program (HST-PID: 13386, PI: Rodney), which aims to identify and study
explosive transients found in the HFF and related
programs\citet{Rodney:2015a}.  The FrontierSN team discovered
\spock\ in two separate HFF observing campaigns on the galaxy cluster
\MACS0416.  The first was an imaging campaign in January, 2014 during
which the MACS0416 cluster field was observed in the F435W, F606W, and
F814W optical bands using the Advanced Camera for Surveys Wide Field
Camera (ACS-WFC).  The second concluded in August, 2014, and imaged
the cluster with the infrared detector of \HST's Wide Field Camera 3
(WFC3-IR) using the F105W, F125W, F140W, and F160W bands.

To discover transient sources, the FrontierSN team processes each new
epoch of \HST data through a difference-imaging
pipeline (\url{https://github.com/srodney/sndrizpipe}), using
archival \HST images to provide reference images (templates) which are
subtracted from the astrometrically registered HFF images. In the case
of MACS0416, the templates were constructed from images collected as
part of the Cluster Lensing And Supernova survey with Hubble (CLASH,
HST-PID:12459, PI:Postman)\cite{Postman:2012}. The resulting difference images are
visually inspected for new point sources, and any new transients of
interest (primarily SNe) are monitored with additional
\HST imaging or ground-based spectroscopic observations as needed.

\subsection{Photometry.}\label{sec:Photometry}

The follow-up observations for \spock\ included \HST imaging
observations in infrared and optical bands using the WFC3-IR and
ACS-WFC detectors, respectively. Tables~\ref{tab:spockonephot} and
\ref{tab:spocktwophot} present photometry of the \spock\ events from
all available \HST observations. The flux was measured on difference
images, first using aperture photometry with a 0\farcs3 radius, and
also by fitting with an empirical point spread function (PSF).  The
PSF model was defined using \HST observations of the G2V standard star
P330E, observed in a separate calibration program.  A separate PSF
model was defined for each filter, but owing to the long-term
stability of the \HST PSF we used the same model in all epochs.  All
of the aperture and PSF-fitting photometry was carried out using the
{\tt PythonPhot} software package
(\url{https://github.com/djones1040/PythonPhot})\citep{Jones:2015}.

\subsection{Host-Galaxy Spectroscopy.}\label{sec:Spectroscopy}

Spectroscopy of the \spock\ host galaxy was collected using three
instruments on the Very Large Telescope (VLT).  Observations with the
VLT's X-shooter cross-dispersed echelle spectrograph\citep{Vernet:2011} were taken on October 19, 21 and 23, 2014
(Program 093.A-0667(A), PI: J. Hjorth) with the slit centered on the
position of \spocktwo.  The total integration time was 4.0 hours for
the NIR arm of X-shooter, 3.6 hours for the VIS arm, and 3.9 hours for
the UVB arm.  The spectrum did not provide any detection of the
transient source itself (as we will see below, it had already faded
back to its quiescent state by that time).  However, it did provide an
unambiguous redshift for the host galaxy of $z=1.0054\pm0.0002$ from
\Ha\ and the \forbidden{O}{ii} doublet in data from the NIR and VIS
arms, respectively.  These line identifications are consistent with
two measures of the photometric redshift of the host: $z=1.00\pm0.02$
from the BPZ algorithm\citep{Benitez:2000}, and $z=0.92\pm0.05$ from
the EAZY program\citep{Brammer:2008}.  Both were derived from \HST
photometry of the host images 11.1 and 11.2, spanning 4350--16000 \AA.

Additional VLT observations were collected using the Visible
Multi-object Spectrograph (VIMOS)\citep{LeFevre:2003}, as part of the
CLASH-VLT large program (Program 186.A-0.798; P.I.:
P. Rosati)\citep{Rosati:2014}, which collected $\sim$4000 reliable
redshifts over 600 arcmin$^2$ in the \macs0416
field\citep{Grillo:2015,Balestra:2016}.  These massively multi-object
observations could potentially have provided confirmation of the
redshift of the \spock host galaxy with separate spectral line
identifications in each of the three host-galaxy images.  For the
\macs0416 field the CLASH-VLT program collected 1 hour of useful
exposure time in good seeing conditions with the Low Resolution Blue
grism.  Unfortunately, the wavelength range of this grism (3600-6700
\AA) does not include any strong emission lines for a source at
$z=1.0054$, and the signal-to-noise ratio (S/N) was not sufficient to provide
any clear line identifications for the three images of the \spock host
galaxy.

The VLT Multi Unit Spectroscopic Explorer (MUSE)\citep{Henault:2003,Bacon:2012} observed the NE portion of
the MACS0416 field---where the \spock host images are located---in
December, 2014 for 2 hours of integration time (ESO program
094.A-0115, PI: J.\,Richard).  These observations also confirmed the
redshift of the host galaxy with clear detection of the
\forbidden{O}{ii} doublet.  Importantly, since MUSE is an integral
field spectrograph, these observations also provided a confirmation of
the redshift of the third image of the host galaxy, 11.3, with a
matching \forbidden{O}{ii} line at the same wavelength\cite{Caminha:2017}.

A final source of spectroscopic information relevant to \spock is the
Grism Lens Amplified Survey from Space (GLASS; PID:
  HST-GO-13459; PI:T. Treu)\citep{Schmidt:2014,Treu:2015a}. The GLASS
program collected slitless spectroscopy on the \macs0416 field using
the WFC3-IR G102 and G141 grisms on \HST, deriving redshifts for
galaxies down to a magnitude limit $H<23$.  As with the VLT VIMOS
data, the three sources identified as images of the \spock host galaxy
are too faint in the GLASS data to provide any useful line
identifications.  There are also no other sources in the GLASS
redshift catalog (\url{http://glass.astro.ucla.edu/}) that
have a spectroscopic redshift consistent with $z=1.0054$.

 \subsection{Gravitational Lens Models.}\label{sec:LensingModels}


The seven lens models used to provide estimates of the plausible range
of magnifications and time delays are as follows:

\begin{itemize}
\item{\it CATS:} The model of \citeref{Jauzac:2014}, version 4.1,
  generated with the {\tt LENSTOOL} software
  (\url{http://projects.lam.fr/repos/lenstool/wiki})\citep{Jullo:2007}
  using strong lensing constraints.  This model parameterizes cluster
  and galaxy components using pseudo-isothermal elliptical mass
  distribution (PIEMD) density profiles\citep{Kassiola:1993,
    Limousin:2007}.
\item{\it GLAFIC:} The model of \citeref{Kawamata:2016}, built using
  the {\tt GLAFIC} software
  (\url{http://www.slac.stanford.edu/~oguri/glafic/})\citep{Oguri:2010b}
  with strong-lensing constraints. This model assumes simply
  parametrized mass distributions, and model parameters are
  constrained using positions of more than 100 multiple images.
\item{\it GLEE:} A new model built using the {\tt GLEE}
  software\citep{Suyu:2010b, Suyu:2012} with the same strong-lensing
  constraints used in \citeref{Caminha:2017}, representing mass
  distributions with simply parameterized mass profiles.
\item{\it GRALE:} A free-form, adaptive grid model developed using
  the GRALE software tool\citep{Liesenborgs:2006, Liesenborgs:2007,
    Mohammed:2014, Sebesta:2016}, which implements a genetic algorithm
  to reconstruct the cluster mass distribution with hundreds to
  thousands of projected Plummer\citet{Plummer:1911} density profiles.
\item{\it SWUnited:} The model of \citeref{Hoag:2016}, built using the
  {\tt SWUnited} modeling method\citep{Bradac:2005, Bradac:2009}, in
  which an adaptive pixelated grid iteratively adapts the mass
  distribution to match both strong- and weak-lensing constraints.
  Time delay predictions are not available for this model.
\item{\it WSLAP+:} Created with the {\tt WSLAP+} software
  (\url{http://www.ifca.unican.es/users/jdiego/LensExplorer})\citep{Sendra:2014}:
  Weak and Strong Lensing Analysis Package plus member galaxies (Note:
  no weak-lensing constraints were used for this \MACS0416 model).
\item{\it ZLTM:} A model with strong- and weak-lensing constraints,
  built using the ``light-traces-mass'' (LTM)
  methodology\citep{Zitrin:2009a, Zitrin:2015}, first presented for
  \MACS0416 in \citeref{Zitrin:2013a}.
\end{itemize}

Early versions of the {\it SWUnited}, {\it CATS}, {\it ZLTM} and {\it
  GRALE} models were originally distributed as part of the Hubble
Frontier Fields lens modeling project
(\url{https://archive.stsci.edu/prepds/frontier/lensmodels/}), in
which models were generated based on data available before the start
of the HFF observations to enable rapid early investigations of lensed
sources. The versions of these models applied here are updated to
incorporate additional lensing constraints.  In all cases the lens
modelers made use of strong-lensing constraints (multiply imaged
systems and arcs) derived from \HST imaging collected as part of the
CLASH program\cite{Postman:2012}). These
models also made use of spectroscopic redshifts in the cluster
field\cite{Mann:2012, Christensen:2012, Grillo:2015, Caminha:2017}.
Input weak-lensing constraints were derived from data collected at the
Subaru Telescope by PI K. Umetsu (in prep) and archival imaging.

 \subsection{X-ray Nondetections.}\label{sec:Xray}

The \MACS0416 field was observed by the \Swift X-Ray Telescope
and UltraViolet/Optical Telescope in April 2013.  No source was
detected near the locations of the \spock events (N. Gehrels, private
communication).  The field was also observed by \Chandra with the
ACIS-I instrument for three separate programs.  On June 7, 2009 it was
observed for GO program 10800770 (PI: H.\,Ebeling).  It was revisited
for GTO program 15800052 (PI: S.\,Murray) on November 20, 2013 and for
GO program 15800858 (PI: C.\, Jones) on June 9, August 31, November
26, and December 17, 2014. These \Chandra images show no evidence for
an x-ray emitting point source near the \spock locations on those
dates (S. Murray, private communication).

The \Chandra observations that were closest in time to the observed
\spock events were those taken in August and November, 2014.  The
August 31 observations were coincident with the observed peak of
rest-frame optical emission for the \spocktwo event (on MJD
56900). The November 26 observations correspond to 44 rest-frame days
after the peak of the \spocktwo event. If the \spock events are
UV/optical nova eruption, then these observations most likely did not
coincide with the nova system's supersoft x-ray phase. For a RN system
the x-ray phase typically initiates after a short delay, and persists
for a span of only a few weeks. For example, the most rapid recurrence
nova known, M31N 2008-12a, has exhibited a supersoft x-ray phase from
6 to 18 days after the peak of the optical emission
\citep{Henze:2015a}.

 \subsection{Light Curve Fitting.}\label{sec:LightCurves}

Due to the rapid decline timescale, no observations were collected for
either event that unambiguously show the declining portion of the
light curve. Therefore, we must make some assumptions for the shape of
the light curve in order to quantify the peak luminosity and the
corresponding timescales for the rise and the decline.  We first
approach this with a simplistic model that is piecewise linear in
magnitude vs time. Supplementary Figure~\ref{fig:LinearLightCurveFits} shows
examples of the resulting fits for the two events.  For each fit we
use only the data collected within 3 days of the brightest observed
magnitude, which allows us to fit a linear rise separately for the
F606W and F814W light curves for \spockone and the F125W and F160W
light curves for \spocktwo. To quantify the covariance between the
true peak brightness, the rise time and the decline timescale, we use
the following procedure:

\begin{enumerate}
\item make an assumption for the date of peak, $t_{\rm pk}$;
\item measure the peak magnitude at $t_{\rm pk}$ from the linear fit
  to the rising light-curve data;
\item assume the source reaches a minimum brightness (maximum
  magnitude) of 30 AB mag at the epoch of first observation after the
  peak;
\item draw a line for the declining light curve between the assumed
  peak and the assumed minimum brightness;
\item use that declining light-curve line to measure the timescale for
  the event to drop by 2 mag, $t_2$;
\item make a new assumption for $t_{\rm pk}$ and repeat.
\end{enumerate}

As shown in Supplementary Figure~\ref{fig:LinearLightCurveFits}, the resulting
piecewise linear fits are simplistic, but nevertheless approximately
capture the observed behavior for both events.  Furthermore, since
this toy model is not physically motivated, it allows us to remain
agnostic for the time being as to the astrophysical source(s) driving
these transients.  From these fits we can see that \spockone most
likely reached a peak magnitude between 25 and 26.5 AB mag in both
F814W and F435W, and had a decline timescale $t_2$ of less than 2 days
in the rest frame. The observations of \spocktwo provide less
stringent constraints, but we see that it had a peak magnitude between
23 and 26.5 AB mag in F160W and exhibited a decline time of less than
seven days.  These fits also illustrate the generic fact that a higher
peak brightness corresponds to a longer rise time and a faster decline
timescale, independent of the specific model used.  Changes to the
arbitrary constraints we placed on these linear fits do not
substantially affect these results.

At any assumed value for the time of peak brightness this linear
interpolation gives an estimate of the peak magnitude. We then convert
that to a luminosity (e.g., $\nu L_\nu$ in erg s$^{-1}$) by first
correcting for the luminosity distance assuming a standard \LCDM
cosmology, and then accounting for an assumed lensing magnification,
$\mu$.  The range of plausible lensing magnifications ($10<\mu<100$)
is derived from the union of our seven independent lens models.  This
results in a grid of possible peak luminosities for each event as a
function of magnification and time of peak.  As we are using linear
light curve fits, the assumed time of peak is equivalent to an
assumption for the decline time, which we quantify as $t_2$, the time
over which the transient declines by 2 magnitudes.

 \subsection{LBV Build-up Timescale and Quiescent Luminosity.}\label{sec:LBVbuildup}

To explore some of the physical implications of an LBV classification
for the two \spock events, we first make a rough estimate of the total
radiated energy, which can be computed using the decline timescale
$t_2$ and the peak luminosity $L_{\rm pk}$:

\begin{equation}
  \label{eqn:Erad}
  E_{\rm rad} = \zeta \t2 \Lpk,
\end{equation}

\noindent where $\zeta$ is a dimensionless factor of order unity that
depends on the precise shape of the light
curve\cite{Smith:2011b}. Note that earlier work\cite{Smith:2011b} has
used $t_{1.5}$ instead of $t_2$, which amounts to a different
light-curve shape term, $\zeta$.  Adopting \Lpk$\approx10^{41}$ erg
s$^{-1}$ and \t2$\approx$1 day (as shown in
Fig.~\ref{fig:PeakLuminosityDeclineTime}), we find that the total
radiated energy is $E_{\rm rad}\approx10^{46}$ erg.  A realistic range
for this estimate would span $10^{44}<E_{\rm rad}<10^{47}$ erg, due to
uncertainties in the magnification, bolometric luminosity correction,
decline time, and light-curve shape. These uncertainties
notwithstanding, our estimate falls well within the range of plausible
values for the total radiated energy of a major LBV outburst.

The ``build-up'' timescale\citep{Smith:2011b} matches the radiative
energy released in an LBV eruption event with the radiative energy
produced during the intervening quiescent phase,

\begin{equation}
  \label{eqn:trad}
t_{\rm rad} = \frac{E_{\rm rad}}{L_{\rm qui}} = \t2 \frac{\xi\Lpk}{L_{\rm qui}},
\end{equation}

\noindent where $L_{\rm qui}$ is the luminosity of the LBV progenitor
star during quiescence.

The \spock events are not resolved as individual stars in their
quiescent phase, so we have no useful constraint on the quiescent
luminosity. Thus, instead of using a measured quiescent luminosity to
estimate the build-up timescale, we assume that $t_{\rm rad}$ for
\spock corresponds to the observed rest-frame lag between the two
events, roughly 120 days (this accounts for both cosmic time dilation
and a gravitational lensing time delay of $\sim$40 days). Adopting
$\Lpk=10^{41}$ erg s$^{-1}$ and $\t2=2$ days (see
Figure~\ref{fig:PeakLuminosityDeclineTime}), we infer that the
quiescent luminosity of the \spock progenitor would be
$L_{\rm qui}\approx10^{39.5}$ erg s$^{-1}$ ($M_V\approx-10$ mag).

 \subsection{RN Light-Curve Comparison.}\label{sec:RNLightCurves}


There are ten known RNe in the Milky Way galaxy, and seven of
these exhibit outbursts that decline rapidly, fading by two magnitudes
in less than ten days\citep{Schaefer:2010}. 
Supplementary Figure~\ref{fig:RecurrentNovaLightCurveComparison}
compares the \spock light curves to a composite light curve (the gray
shaded region), which encompasses the V band light curve
templates\citep{Schaefer:2010} for all seven of these galactic RN
events.  The Andromeda galaxy (M31) also hosts at least one RN with a
rapidly declining light curve.  The 2014 eruption of this well-studied
nova, M31N 2008-12a, is shown as a solid black line in Supplementary
Figure~\ref{fig:RecurrentNovaLightCurveComparison}, fading by 2 mag in
$<3$ days.  This comparison demonstrates that the rapid decline of
both of the \spock transient events is fully consistent with the
eruptions of known RNe in the local universe.


\subsection{RN Luminosity and Recurrence Period.}\label{sec:RNLuminosityRecurrence}

To examine the recurrence period and peak brightness of the \spock
events relative to RNe, we rely on a pair of papers that evaluated an
extensive grid of nova models through multiple cycles of outburst and
quiescence\citep{Prialnik:1995,Yaron:2005}.  Supplementary
Figure~\ref{fig:RecurrentNovaRecurrenceComparison} plots first the RN
outburst amplitude (the apparent magnitude between outbursts minus the
apparent magnitude at peak) and then the peak luminosity against the
log of the recurrence period in years.
For the \spock events we can only measure a lower limit on the
outburst amplitude, since the presumed progenitor star is unresolved,
so no measurement is available at quiescence. Supplementary
Figure~\ref{fig:RecurrentNovaRecurrenceComparison} shows that a
recurrence period as fast as one year is expected only for a RN system
in which the primary white dwarf is both very close to the
Chandrasekhar mass limit (1.4 \Msun) and also has an extraordinarily
rapid mass transfer rate ($\sim10^{-6}$ \Msun yr$^{-1}$).  The models
of \citeref{Yaron:2005} suggest that such systems should have a very
low peak amplitude (barely consistent with the lower limit for \spock)
and a low peak luminosity ($\sim$100 times less luminous than the
\spock events).

The closest analog for the \spock events from the population of known
RN systems is the nova M31N\,2008-12a.  \citeref{Kato:2015} provided a
theoretical model that can account for the key observational
characteristics of this remarkable nova: the very rapid recurrence
timescale ($<$1 yr), fast optical light curve ($\t2\sim2$ days), and
short supersoft x-ray phase (6-18 days after optical
  outburst)\citep{Henze:2015a}.  To match these observations,
\citeref{Kato:2015} invoke a 1.38 \Msun white dwarf primary,
drawing mass from a companion at a rate of $1.6\times10^{-7}$ \Msun
yr$^{-1}$.  This is largely consistent with the theoretical
expectations derived by \citeref{Yaron:2005}, and reinforces the
conclusion that a combination of a high-mass white dwarf and efficient
mass transfer are the key ingredients for rapid recurrence and short
light curves. The one feature that cannot be effectively explained
with this hypothesis is the peculiarly high luminosity of the \spock
events -- even after accounting for the very large uncertainties. 

 \subsection{Intracluster Light.}\label{sec:ICL}

To estimate the mass of intracluster stars along the line of sight to
the \spock events, we follow the procedure of \citeref{Kelly:2017} and
Morishita et al. (in prep).  This entails fitting and removing the
surface brightness of individual galaxies in the field, then fitting a
smooth profile to the residual surface brightness of intracluster
light (ICL).  The surface brightness is then converted to a projected
stellar mass surface density by assuming a
Chabrier\cite{Chabrier:2003} initial mass function and an
exponentially declining star formation history.  This procedure leads
to an estimate for the intracluster stellar mass of $\log
(\Sigma_{\star} / (\Msun\,{\rm kpc}^{-2})) = 6.9\pm0.4$.  This is
very similar to the value of $6.8^{+0.4}_{-0.3}$ inferred for the
probable caustic crossing star M1149 LS1\cite{Kelly:2017}.

 \subsection{Color Curves.}\label{sec:ColorCurves}

At $z=1$ the observed optical and infrared bands translate to
rest-frame ultraviolet (UV) and optical wavelengths, respectively.  To
derive rest-frame UV and optical colors from the observed photometry,
we start with the measured magnitude in a relatively blue band (F435W
and F606W for \spockone and F105W, F125W, F140W for \spocktwo).  We
then subtract the coeval magnitude for a matched red band (F814W for
\spockone, F125W or F160W for \spocktwo), derived from the linear fits
to those bands.  To adjust these to rest-frame filters, we apply K
corrections\citep{Hogg:2002}, which we compute by
defining a crude SED via linear interpolation between the observed
broad bands for each transient event at each epoch.  For consistency
with past published results, we include in each K correction a
transformation from AB to Vega-based magnitudes.  The resulting UV and
optical colors are plotted in Supplementary Figure~\ref{fig:ColorCurves}.  Both
\spockone and \spocktwo show little or no color variation over the
period where color information is available.  This lack of color
evolution is compatible with all three of the primary hypotheses
advanced, as it is possible to have no discernible color evolution
from either an LBV or RN over this short time span, and microlensing
events inherently exhibit an unchanging color.

If these two events are from a single source then one could construct
a composite SED from rest-frame UV to optical wavelengths by combining
the NW and SE flux measurements, but only after correcting for the
relative magnification.  Figure~\ref{fig:LightCurves}
shows that the observed peak brightnesses for the two events agree to
within $\sim30\%$.  This implies that for any composite SED, the
rest-frame UV to optical flux ratio is approximately equal to the
NW:SE magnification ratio, and any extreme asymmetry in the
magnification would indicate a very steep slope in the SED.

 \subsection{Rates.}\label{sec:RatesMethods}

To derive a rough estimate of the rate of \spock-like transients, we
first define the set of strongly lensed galaxies in which a similarly
faint and fast transient could have been detected in the HFF
imaging. The single-epoch detection limit of the HFF transient search
was $m_{\rm lim}=26.7$ AB mag, consistent with the SN searches carried
out in the CLASH and CANDELS programs\cite{Graur:2014,Rodney:2014}.
For a transient with peak brightness $M_{V}>-14$ mag to be detected,
the host galaxy must be amplified by strong lensing with a
magnification $\mu>20$ at $z\sim1$, growing to $\mu>100$ at $z\sim2$.
Using photometric redshifts and magnifications derived from the GLAFIC
lens models of the six HFF clusters, we find $N_{\rm gal}=6$ galaxies
that satisfy this criterion, with $0.5<z<1.5$ (Supplementary
Figure~\ref{fig:StronglyLensedGalaxies}).

We then define the {\it control time}, $t_{c}$, for the HFF survey,
which gives the span of time over which each cluster was observed with
a cadence sufficient for detection of such rapid transients.  We
define this as any period in which at least two \HST observations were
collected within every 10 day span. This effectively includes the
entirety of the primary HFF campaigns on each cluster, but excludes
all of the ancillary data collection periods from supplemental \HST
imaging programs. The average control time for an HFF cluster is
$t_{c} = 0.22$ years (80 days).  Treating each \spock event as a
separate detection, we can derive a rate estimate using $R = 2 /
(N_{\rm gal}\,t_c)$.  This yields $R=1.5$ events galaxy$^{-1}$
year$^{-1}$.   

Future examination of the rate of such transients should consider the
total stellar mass and the star-formation rates of the galaxies
surveyed, or use a projection of the lensed background area onto the
source plane to derive a volumetric rate.  Such analyses would require
a more detailed exploration of the impact of lensing uncertainties on
derived properties of the lensed galaxies and the lensed volume, and
this is beyond the scope of the current work.

\end{methods}


\begin{addendum}
 \item[Supplementary Information] Supplementary figures, tables and notes are included at the end of this document.
 \item The authors thank Mario Livio and Laura Chomiuk for helpful discussion
of this paper, as well as Stephen Murray and Neil Gehrels for
assistance with the \Chandra and \Swift data, respectively.

Financial support for this work was provided to S.A.R., O.G., and L.G.S. by NASA
through grant HST-GO-13386 from the Space Telescope Science Institute
(STScI), which is operated by Associated Universities for Research in
Astronomy, Inc. (AURA), under NASA contract NAS 5-26555.
J.M.D acknowledges support of the
projects AYA2015-64508-P (MINECO/FEDER, UE), AYA2012-39475-C02-01 and
the consolider project CSD2010-00064 funded by the Ministerio de
Economia y Competitividad.
A.V.F. and P.L.K. are grateful for financial assistance from the
Christopher R. Redlich Fund, the TABASGO Foundation, and NASA/STScI
grants 14528, 14872, and 14922.  The work of A.V.F. was conducted in
part at the Aspen Center for Physics, which is supported by NSF grant
PHY-1607611; he thanks the Center for its hospitality during the
neutron stars workshop in June and July 2017.
R.J.F. and the UCSC group is supported in part by NSF grant
AST-1518052 and from fellowships from the Alfred P.\ Sloan Foundation
and the David and Lucile Packard Foundation to R.J.F.
C.G. acknowledges support by VILLUM FONDEN Young Investigator Programme through grant no. 10123.
M.J. was supported by the Science and
Technology Facilities Council (grant number ST/L00075X/1) and used the
DiRAC Data Centric system at Durham University, operated by the
Institute for Computational Cosmology on behalf of the STFC DiRAC HPC
Facility (\url{www.dirac.ac.uk}).  M.J. was funded by BIS National
E-infrastructure capital grant ST/K00042X/1, STFC capital grant
ST/H008519/1, and STFC DiRAC Operations grant ST/K003267/1 and Durham
University. DiRAC is part of the National E-Infrastructure.
R.K. was supported by Grant-in-Aid for JSPS Research Fellow (16J01302).
M.O.  acknowledges support in part by World Premier International
Research Center Initiative (WPI Initiative), MEXT, Japan, and JSPS
KAKENHI Grant Number 26800093 and 15H05892.
J.R. acknowledges support from the ERC starting grant
336736-CALENDS.
G.C. and S.H.S. thank the Max Planck Society for support through the Max Planck
Research Group of S.H.S.
T.T. and the GLASS team were funded by NASA through HST grant
HST-GO-13459 from STScI.
L.L.R.W. would like to thank Minnesota Supercomputing Institute at
the University of Minnesota for providing resources and support.

 \item[Correspondence] Correspondence and requests for materials
should be addressed to S.A.R.~(email: srodney@sc.edu).
\end{addendum}



\begin{thebibliography}{10}
\expandafter\ifx\csname url\endcsname\relax
  \def\url#1{\texttt{#1}}\fi
\expandafter\ifx\csname urlprefix\endcsname\relax\def\urlprefix{URL }\fi
\providecommand{\bibinfo}[2]{#2}
\providecommand{\eprint}[2][]{\url{#2}}

\bibitem{Kasliwal:2011a}
\bibinfo{author}{{Kasliwal}, M.~M.} \emph{et~al.}
\newblock \bibinfo{title}{{Discovery of a New Photometric Sub-class of Faint
  and Fast Classical Novae}}.
\newblock \emph{\bibinfo{journal}{\apj}} \textbf{\bibinfo{volume}{735}},
  \bibinfo{pages}{94} (\bibinfo{year}{2011}).

\bibitem{Drout:2014}
\bibinfo{author}{{Drout}, M.~R.} \emph{et~al.}
\newblock \bibinfo{title}{{Rapidly Evolving and Luminous Transients from
  Pan-STARRS1}}.
\newblock \emph{\bibinfo{journal}{\apj}} \textbf{\bibinfo{volume}{794}},
  \bibinfo{pages}{23} (\bibinfo{year}{2014}).

\bibitem{Berger:2013b}
\bibinfo{author}{{Berger}, E.} \emph{et~al.}
\newblock \bibinfo{title}{{A Search for Fast Optical Transients in the
  Pan-STARRS1 Medium-Deep Survey: M-Dwarf Flares, Asteroids, Limits on
  Extragalactic Rates, and Implications for LSST}}.
\newblock \emph{\bibinfo{journal}{\apj}} \textbf{\bibinfo{volume}{779}},
  \bibinfo{pages}{18} (\bibinfo{year}{2013}).

\bibitem{Tyson:2002}
\bibinfo{author}{{Tyson}, J.~A.}
\newblock \bibinfo{title}{{Large Synoptic Survey Telescope: Overview}}.
\newblock In \bibinfo{editor}{{Tyson}, J.~A.} \& \bibinfo{editor}{{Wolff}, S.}
  (eds.) \emph{\bibinfo{booktitle}{Survey and Other Telescope Technologies and
  Discoveries}}, vol. \bibinfo{volume}{4836} of \emph{\bibinfo{series}{Society
  of Photo-Optical Instrumentation Engineers (SPIE) Conference Series}},
  \bibinfo{pages}{10--20} (\bibinfo{year}{2002}).

\bibitem{Lotz:2017}
\bibinfo{author}{{Lotz}, J.~M.} \emph{et~al.}
\newblock \bibinfo{title}{{The Frontier Fields: Survey Design and Initial
  Results}}.
\newblock \emph{\bibinfo{journal}{\apj}} \textbf{\bibinfo{volume}{837}},
  \bibinfo{pages}{97} (\bibinfo{year}{2017}).

\bibitem{Caminha:2017}
\bibinfo{author}{{Caminha}, G.~B.} \emph{et~al.}
\newblock \bibinfo{title}{{A refined mass distribution of the cluster MACS
  J0416.1-2403 from a new large set of spectroscopic multiply lensed sources}}.
\newblock \emph{\bibinfo{journal}{\aap}} \textbf{\bibinfo{volume}{600}},
  \bibinfo{pages}{A90} (\bibinfo{year}{2017}).

\bibitem{Pastorello:2013}
\bibinfo{author}{{Pastorello}, A.} \emph{et~al.}
\newblock \bibinfo{title}{{Interacting Supernovae and Supernova Impostors: SN
  2009ip, is this the End?}}
\newblock \emph{\bibinfo{journal}{\apj}} \textbf{\bibinfo{volume}{767}},
  \bibinfo{pages}{1} (\bibinfo{year}{2013}).

\bibitem{Pastorello:2010}
\bibinfo{author}{{Pastorello}, A.} \emph{et~al.}
\newblock \bibinfo{title}{{Multiple major outbursts from a restless luminous
  blue variable in NGC 3432}}.
\newblock \emph{\bibinfo{journal}{\mnras}} \textbf{\bibinfo{volume}{408}},
  \bibinfo{pages}{181--198} (\bibinfo{year}{2010}).

\bibitem{Zitrin:2013a}
\bibinfo{author}{{Zitrin}, A.} \emph{et~al.}
\newblock \bibinfo{title}{{CLASH: The Enhanced Lensing Efficiency of the Highly
  Elongated Merging Cluster MACS J0416.1-2403}}.
\newblock \emph{\bibinfo{journal}{\apjl}} \textbf{\bibinfo{volume}{762}},
  \bibinfo{pages}{L30} (\bibinfo{year}{2013}).

\bibitem{Jauzac:2014}
\bibinfo{author}{{Jauzac}, M.} \emph{et~al.}
\newblock \bibinfo{title}{{Hubble Frontier Fields: a high-precision
  strong-lensing analysis of galaxy cluster MACSJ0416.1-2403 using 200 multiple
  images}}.
\newblock \emph{\bibinfo{journal}{\mnras}} \textbf{\bibinfo{volume}{443}},
  \bibinfo{pages}{1549--1554} (\bibinfo{year}{2014}).

\bibitem{Johnson:2014}
\bibinfo{author}{{Johnson}, T.~L.} \emph{et~al.}
\newblock \bibinfo{title}{{Lens Models and Magnification Maps of the Six Hubble
  Frontier Fields Clusters}}.
\newblock \emph{\bibinfo{journal}{\apj}} \textbf{\bibinfo{volume}{797}},
  \bibinfo{pages}{48} (\bibinfo{year}{2014}).

\bibitem{Richard:2014}
\bibinfo{author}{{Richard}, J.} \emph{et~al.}
\newblock \bibinfo{title}{{Mass and magnification maps for the Hubble Space
  Telescope Frontier Fields clusters: implications for high-redshift studies}}.
\newblock \emph{\bibinfo{journal}{\mnras}} \textbf{\bibinfo{volume}{444}},
  \bibinfo{pages}{268--289} (\bibinfo{year}{2014}).

\bibitem{Diego:2015a}
\bibinfo{author}{{Diego}, J.~M.} \emph{et~al.}
\newblock \bibinfo{title}{{A free-form lensing grid solution for A1689 with new
  multiple images}}.
\newblock \emph{\bibinfo{journal}{\mnras}} \textbf{\bibinfo{volume}{446}},
  \bibinfo{pages}{683--704} (\bibinfo{year}{2015}).

\bibitem{Grillo:2015}
\bibinfo{author}{{Grillo}, C.} \emph{et~al.}
\newblock \bibinfo{title}{{CLASH-VLT: Insights on the Mass Substructures in the
  Frontier Fields Cluster MACS J0416.1-2403 through Accurate Strong Lens
  Modeling}}.
\newblock \emph{\bibinfo{journal}{\apj}} \textbf{\bibinfo{volume}{800}},
  \bibinfo{pages}{38} (\bibinfo{year}{2015}).

\bibitem{Hoag:2016}
\bibinfo{author}{{Hoag}, A.} \emph{et~al.}
\newblock \bibinfo{title}{{The Grism Lens-Amplified Survey from Space (GLASS).
  VI. Comparing the Mass and Light in MACS J0416.1-2403 Using Frontier Field
  Imaging and GLASS Spectroscopy}}.
\newblock \emph{\bibinfo{journal}{\apj}} \textbf{\bibinfo{volume}{831}},
  \bibinfo{pages}{182} (\bibinfo{year}{2016}).

\bibitem{Sebesta:2016}
\bibinfo{author}{{Sebesta}, K.}, \bibinfo{author}{{Williams}, L.~L.~R.},
  \bibinfo{author}{{Mohammed}, I.}, \bibinfo{author}{{Saha}, P.} \&
  \bibinfo{author}{{Liesenborgs}, J.}
\newblock \bibinfo{title}{{Testing light-traces-mass in Hubble Frontier Fields
  Cluster MACS-J0416.1-2403}}.
\newblock \emph{\bibinfo{journal}{\mnras}} \textbf{\bibinfo{volume}{461}},
  \bibinfo{pages}{2126--2134} (\bibinfo{year}{2016}).

\bibitem{Karoff:2016}
\bibinfo{author}{{Karoff}, C.} \emph{et~al.}
\newblock \bibinfo{title}{{Observational evidence for enhanced magnetic
  activity of superflare stars}}.
\newblock \emph{\bibinfo{journal}{Nature Communications}}
  \textbf{\bibinfo{volume}{7}}, \bibinfo{pages}{11058} (\bibinfo{year}{2016}).

\bibitem{Fregeau:2004}
\bibinfo{author}{{Fregeau}, J.~M.}, \bibinfo{author}{{Cheung}, P.},
  \bibinfo{author}{{Portegies Zwart}, S.~F.} \& \bibinfo{author}{{Rasio},
  F.~A.}
\newblock \bibinfo{title}{{Stellar collisions during binary-binary and
  binary-single star interactions}}.
\newblock \emph{\bibinfo{journal}{\mnras}} \textbf{\bibinfo{volume}{352}},
  \bibinfo{pages}{1--19} (\bibinfo{year}{2004}).

\bibitem{Metzger:2012}
\bibinfo{author}{{Metzger}, B.~D.}, \bibinfo{author}{{Giannios}, D.} \&
  \bibinfo{author}{{Spiegel}, D.~S.}
\newblock \bibinfo{title}{{Optical and X-ray transients from planet-star
  mergers}}.
\newblock \emph{\bibinfo{journal}{\mnras}} \textbf{\bibinfo{volume}{425}},
  \bibinfo{pages}{2778--2798} (\bibinfo{year}{2012}).

\bibitem{Yamazaki:2017}
\bibinfo{author}{{Yamazaki}, R.}, \bibinfo{author}{{Hayasaki}, K.} \&
  \bibinfo{author}{{Loeb}, A.}
\newblock \bibinfo{title}{{Optical-infrared flares and radio afterglows by
  Jovian planets inspiraling into their host stars}}.
\newblock \emph{\bibinfo{journal}{\mnras}} \textbf{\bibinfo{volume}{466}},
  \bibinfo{pages}{1421--1427} (\bibinfo{year}{2017}).

\bibitem{Di-Stefano:2015}
\bibinfo{author}{{Di Stefano}, R.}, \bibinfo{author}{{Fisher}, R.},
  \bibinfo{author}{{Guillochon}, J.} \& \bibinfo{author}{{Steiner}, J.~F.}
\newblock \bibinfo{title}{{Death by Dynamics: Planetoid-Induced Explosions on
  White Dwarfs}}.
\newblock \emph{\bibinfo{journal}{ArXiv e-prints}}  (\bibinfo{year}{2015}).

\bibitem{Kulkarni:2007}
\bibinfo{author}{{Kulkarni}, S.~R.} \emph{et~al.}
\newblock \bibinfo{title}{{An unusually brilliant transient in the galaxy
  M85}}.
\newblock \emph{\bibinfo{journal}{\nat}} \textbf{\bibinfo{volume}{447}},
  \bibinfo{pages}{458--460} (\bibinfo{year}{2007}).

\bibitem{Gal-Yam:2012}
\bibinfo{author}{{Gal-Yam}, A.}
\newblock \bibinfo{title}{{Luminous Supernovae}}.
\newblock \emph{\bibinfo{journal}{Science}} \textbf{\bibinfo{volume}{337}},
  \bibinfo{pages}{927--} (\bibinfo{year}{2012}).

\bibitem{Foley:2013a}
\bibinfo{author}{{Foley}, R.~J.} \emph{et~al.}
\newblock \bibinfo{title}{{Type Iax Supernovae: A New Class of Stellar
  Explosion}}.
\newblock \emph{\bibinfo{journal}{\apj}} \textbf{\bibinfo{volume}{767}},
  \bibinfo{pages}{57} (\bibinfo{year}{2013}).

\bibitem{Kasliwal:2012}
\bibinfo{author}{{Kasliwal}, M.~M.} \emph{et~al.}
\newblock \bibinfo{title}{{Calcium-rich Gap Transients in the Remote Outskirts
  of Galaxies}}.
\newblock \emph{\bibinfo{journal}{\apj}} \textbf{\bibinfo{volume}{755}},
  \bibinfo{pages}{161} (\bibinfo{year}{2012}).

\bibitem{Li:1998}
\bibinfo{author}{{Li}, L.-X.} \& \bibinfo{author}{{Paczy{\'n}ski}, B.}
\newblock \bibinfo{title}{{Transient Events from Neutron Star Mergers}}.
\newblock \emph{\bibinfo{journal}{\apjl}} \textbf{\bibinfo{volume}{507}},
  \bibinfo{pages}{L59--L62} (\bibinfo{year}{1998}).

\bibitem{Tanvir:2013}
\bibinfo{author}{{Tanvir}, N.~R.} \emph{et~al.}
\newblock \bibinfo{title}{{A `kilonova' associated with the short-duration
  {$\gamma$}-ray burst GRB 130603B}}.
\newblock \emph{\bibinfo{journal}{\nat}} \textbf{\bibinfo{volume}{500}},
  \bibinfo{pages}{547--549} (\bibinfo{year}{2013}).

\bibitem{Jin:2016}
\bibinfo{author}{{Jin}, Z.-P.} \emph{et~al.}
\newblock \bibinfo{title}{{The Macronova in GRB 050709 and the GRB-macronova
  connection}}.
\newblock \emph{\bibinfo{journal}{Nature Communications}}
  \textbf{\bibinfo{volume}{7}}, \bibinfo{pages}{12898} (\bibinfo{year}{2016}).

\bibitem{Bildsten:2007}
\bibinfo{author}{{Bildsten}, L.}, \bibinfo{author}{{Shen}, K.~J.},
  \bibinfo{author}{{Weinberg}, N.~N.} \& \bibinfo{author}{{Nelemans}, G.}
\newblock \bibinfo{title}{{Faint Thermonuclear Supernovae from AM Canum
  Venaticorum Binaries}}.
\newblock \emph{\bibinfo{journal}{\apjl}} \textbf{\bibinfo{volume}{662}},
  \bibinfo{pages}{L95--L98} (\bibinfo{year}{2007}).

\bibitem{Barnes:2013}
\bibinfo{author}{{Barnes}, J.} \& \bibinfo{author}{{Kasen}, D.}
\newblock \bibinfo{title}{{Effect of a High Opacity on the Light Curves of
  Radioactively Powered Transients from Compact Object Mergers}}.
\newblock \emph{\bibinfo{journal}{\apj}} \textbf{\bibinfo{volume}{775}},
  \bibinfo{pages}{18} (\bibinfo{year}{2013}).

\bibitem{Kasen:2015}
\bibinfo{author}{{Kasen}, D.}, \bibinfo{author}{{Fern{\'a}ndez}, R.} \&
  \bibinfo{author}{{Metzger}, B.~D.}
\newblock \bibinfo{title}{{Kilonova light curves from the disc wind outflows of
  compact object mergers}}.
\newblock \emph{\bibinfo{journal}{\mnras}} \textbf{\bibinfo{volume}{450}},
  \bibinfo{pages}{1777--1786} (\bibinfo{year}{2015}).

\bibitem{Shen:2010}
\bibinfo{author}{{Shen}, K.~J.}, \bibinfo{author}{{Kasen}, D.},
  \bibinfo{author}{{Weinberg}, N.~N.}, \bibinfo{author}{{Bildsten}, L.} \&
  \bibinfo{author}{{Scannapieco}, E.}
\newblock \bibinfo{title}{{Thermonuclear .Ia Supernovae from Helium Shell
  Detonations: Explosion Models and Observables}}.
\newblock \emph{\bibinfo{journal}{\apj}} \textbf{\bibinfo{volume}{715}},
  \bibinfo{pages}{767--774} (\bibinfo{year}{2010}).

\bibitem{Smith:2011b}
\bibinfo{author}{{Smith}, N.}, \bibinfo{author}{{Li}, W.},
  \bibinfo{author}{{Silverman}, J.~M.}, \bibinfo{author}{{Ganeshalingam}, M.}
  \& \bibinfo{author}{{Filippenko}, A.~V.}
\newblock \bibinfo{title}{{Luminous blue variable eruptions and related
  transients: diversity of progenitors and outburst properties}}.
\newblock \emph{\bibinfo{journal}{\mnras}} \textbf{\bibinfo{volume}{415}},
  \bibinfo{pages}{773--810} (\bibinfo{year}{2011}).

\bibitem{Kochanek:2012}
\bibinfo{author}{{Kochanek}, C.~S.}, \bibinfo{author}{{Szczygie{\l}}, D.~M.} \&
  \bibinfo{author}{{Stanek}, K.~Z.}
\newblock \bibinfo{title}{{Unmasking the Supernova Impostors}}.
\newblock \emph{\bibinfo{journal}{\apj}} \textbf{\bibinfo{volume}{758}},
  \bibinfo{pages}{142} (\bibinfo{year}{2012}).

\bibitem{Maza:2009}
\bibinfo{author}{{Maza}, J.} \emph{et~al.}
\newblock \bibinfo{title}{Supernova 2009ip in ngc 7259}.
\newblock \emph{\bibinfo{journal}{CBET}} \bibinfo{pages}{1}
  (\bibinfo{year}{2009}).

\bibitem{Kilpatrick:2017}
\bibinfo{author}{{Kilpatrick}, C.~D.} \emph{et~al.}
\newblock \bibinfo{title}{{Connecting the progenitors, pre-explosion
  variability, and giant outbursts of luminous blue variables with Gaia16cfr}}.
\newblock \emph{\bibinfo{journal}{arXiv: 1706.09962}}  (\bibinfo{year}{2017}).

\bibitem{Pastorello:2017}
\bibinfo{author}{{Pastorello}, A.} \emph{et~al.}
\newblock \bibinfo{title}{{Supernovae 2016bdu and 2005gl, and their link with
  SN 2009ip-like transients: another piece of the puzzle}}.
\newblock \emph{\bibinfo{journal}{arXiv: 1707.00611}}  (\bibinfo{year}{2017}).

\bibitem{DellaValle:1995}
\bibinfo{author}{{Della Valle}, M.} \& \bibinfo{author}{{Livio}, M.}
\newblock \bibinfo{title}{{The Calibration of Novae as Distance Indicators}}.
\newblock \emph{\bibinfo{journal}{\apj}} \textbf{\bibinfo{volume}{452}},
  \bibinfo{pages}{704} (\bibinfo{year}{1995}).

\bibitem{Downes:2000}
\bibinfo{author}{{Downes}, R.~A.} \& \bibinfo{author}{{Duerbeck}, H.~W.}
\newblock \bibinfo{title}{{Optical Imaging of Nova Shells and the Maximum
  Magnitude-Rate of Decline Relationship}}.
\newblock \emph{\bibinfo{journal}{\aj}} \textbf{\bibinfo{volume}{120}},
  \bibinfo{pages}{2007--2037} (\bibinfo{year}{2000}).

\bibitem{Schaefer:2010}
\bibinfo{author}{{Schaefer}, B.~E.}
\newblock \bibinfo{title}{{Comprehensive Photometric Histories of All Known
  Galactic Recurrent Novae}}.
\newblock \emph{\bibinfo{journal}{\apjs}} \textbf{\bibinfo{volume}{187}},
  \bibinfo{pages}{275--373} (\bibinfo{year}{2010}).

\bibitem{Tang:2014}
\bibinfo{author}{{Tang}, S.} \emph{et~al.}
\newblock \bibinfo{title}{{An Accreting White Dwarf near the Chandrasekhar
  Limit in the Andromeda Galaxy}}.
\newblock \emph{\bibinfo{journal}{\apj}} \textbf{\bibinfo{volume}{786}},
  \bibinfo{pages}{61} (\bibinfo{year}{2014}).

\bibitem{Darnley:2016}
\bibinfo{author}{{Darnley}, M.~J.} \emph{et~al.}
\newblock \bibinfo{title}{{M31N 2008-12a - The Remarkable Recurrent Nova in
  M31: Panchromatic Observations of the 2015 Eruption.}}
\newblock \emph{\bibinfo{journal}{\apj}} \textbf{\bibinfo{volume}{833}},
  \bibinfo{pages}{149} (\bibinfo{year}{2016}).

\bibitem{Yaron:2005}
\bibinfo{author}{{Yaron}, O.}, \bibinfo{author}{{Prialnik}, D.},
  \bibinfo{author}{{Shara}, M.~M.} \& \bibinfo{author}{{Kovetz}, A.}
\newblock \bibinfo{title}{{An Extended Grid of Nova Models. II. The Parameter
  Space of Nova Outbursts}}.
\newblock \emph{\bibinfo{journal}{\apj}} \textbf{\bibinfo{volume}{623}},
  \bibinfo{pages}{398--410} (\bibinfo{year}{2005}).

\bibitem{Schneider:1986}
\bibinfo{author}{{Schneider}, P.} \& \bibinfo{author}{{Weiss}, A.}
\newblock \bibinfo{title}{{The two-point-mass lens - Detailed investigation of
  a special asymmetric gravitational lens}}.
\newblock \emph{\bibinfo{journal}{\aap}} \textbf{\bibinfo{volume}{164}},
  \bibinfo{pages}{237--259} (\bibinfo{year}{1986}).

\bibitem{MiraldaEscude:1991}
\bibinfo{author}{{MiraldaEscude}, J.}
\newblock \bibinfo{title}{{The magnification of stars crossing a caustic. I -
  Lenses with smooth potentials}}.
\newblock \emph{\bibinfo{journal}{\apj}} \textbf{\bibinfo{volume}{379}},
  \bibinfo{pages}{94--98} (\bibinfo{year}{1991}).

\bibitem{Lewis:1993}
\bibinfo{author}{{Lewis}, G.~F.}, \bibinfo{author}{{Miralda-Escude}, J.},
  \bibinfo{author}{{Richardson}, D.~C.} \& \bibinfo{author}{{Wambsganss}, J.}
\newblock \bibinfo{title}{{Microlensing light curves - A new and efficient
  numerical method}}.
\newblock \emph{\bibinfo{journal}{\mnras}} \textbf{\bibinfo{volume}{261}},
  \bibinfo{pages}{647--656} (\bibinfo{year}{1993}).

\bibitem{Diego:2017}
\bibinfo{author}{{Diego}, J.~M.} \emph{et~al.}
\newblock \bibinfo{title}{{Dark matter under the microscope: Constraining
  compact dark matter with caustic crossing events}}.
\newblock \emph{\bibinfo{journal}{arXiv:1706.10281}}  (\bibinfo{year}{2017}).

\bibitem{Kelly:2017}
\bibinfo{author}{{Kelly}, P.~L.} \emph{et~al.}
\newblock \bibinfo{title}{{An individual star at redshift 1.5 extremely
  magnified by a galaxy-cluster lens}}.
\newblock \emph{\bibinfo{journal}{arXiv:1706.10279}}  (\bibinfo{year}{2017}).

\bibitem{Rodney:2015a}
\bibinfo{author}{{Rodney}, S.~A.} \emph{et~al.}
\newblock \bibinfo{title}{{Illuminating a Dark Lens : A Type Ia Supernova
  Magnified by the Frontier Fields Galaxy Cluster Abell 2744}}.
\newblock \emph{\bibinfo{journal}{\apj}} \textbf{\bibinfo{volume}{811}},
  \bibinfo{pages}{70} (\bibinfo{year}{2015}).

\bibitem{Postman:2012}
\bibinfo{author}{{Postman}, M.} \emph{et~al.}
\newblock \bibinfo{title}{{The Cluster Lensing and Supernova Survey with
  Hubble: An Overview}}.
\newblock \emph{\bibinfo{journal}{\apjs}} \textbf{\bibinfo{volume}{199}},
  \bibinfo{pages}{25} (\bibinfo{year}{2012}).

\bibitem{Jones:2015}
\bibinfo{author}{{Jones}, D.~O.}, \bibinfo{author}{{Scolnic}, D.~M.} \&
  \bibinfo{author}{{Rodney}, S.~A.}
\newblock \bibinfo{title}{{PythonPhot: Simple DAOPHOT-type photometry in
  Python}}.
\newblock \bibinfo{howpublished}{Astrophysics Source Code Library}
  (\bibinfo{year}{2015}).

\bibitem{Vernet:2011}
\bibinfo{author}{{Vernet}, J.} \emph{et~al.}
\newblock \bibinfo{title}{{X-shooter, the new wide band intermediate resolution
  spectrograph at the ESO Very Large Telescope}}.
\newblock \emph{\bibinfo{journal}{\aap}} \textbf{\bibinfo{volume}{536}},
  \bibinfo{pages}{A105} (\bibinfo{year}{2011}).

\bibitem{Benitez:2000}
\bibinfo{author}{{Ben{\'{\i}}tez}, N.}
\newblock \bibinfo{title}{{Bayesian Photometric Redshift Estimation}}.
\newblock \emph{\bibinfo{journal}{\apj}} \textbf{\bibinfo{volume}{536}},
  \bibinfo{pages}{571--583} (\bibinfo{year}{2000}).

\bibitem{Brammer:2008}
\bibinfo{author}{{Brammer}, G.~B.}, \bibinfo{author}{{van Dokkum}, P.~G.} \&
  \bibinfo{author}{{Coppi}, P.}
\newblock \bibinfo{title}{{EAZY: A Fast, Public Photometric Redshift Code}}.
\newblock \emph{\bibinfo{journal}{\apj}} \textbf{\bibinfo{volume}{686}},
  \bibinfo{pages}{1503--1513} (\bibinfo{year}{2008}).

\bibitem{LeFevre:2003}
\bibinfo{author}{{Le F{\`e}vre}, O.} \emph{et~al.}
\newblock \bibinfo{title}{{Commissioning and performances of the VLT-VIMOS
  instrument}}.
\newblock In \bibinfo{editor}{{Iye}, M.} \& \bibinfo{editor}{{Moorwood},
  A.~F.~M.} (eds.) \emph{\bibinfo{booktitle}{Instrument Design and Performance
  for Optical/Infrared Ground-based Telescopes}}, vol. \bibinfo{volume}{4841}
  of \emph{\bibinfo{series}{Society of Photo-Optical Instrumentation Engineers
  (SPIE) Conference Series}}, \bibinfo{pages}{1670--1681}
  (\bibinfo{year}{2003}).

\bibitem{Rosati:2014}
\bibinfo{author}{{Rosati}, P.} \emph{et~al.}
\newblock \bibinfo{title}{{CLASH-VLT: A VIMOS Large Programme to Map the Dark
  Matter Mass Distribution in Galaxy Clusters and Probe Distant Lensed
  Galaxies}}.
\newblock \emph{\bibinfo{journal}{The Messenger}}
  \textbf{\bibinfo{volume}{158}}, \bibinfo{pages}{48--53}
  (\bibinfo{year}{2014}).

\bibitem{Balestra:2016}
\bibinfo{author}{{Balestra}, I.} \emph{et~al.}
\newblock \bibinfo{title}{{CLASH-VLT: Dissecting the Frontier Fields Galaxy
  Cluster MACS J0416.1-2403 with 800 Spectra of Member Galaxies}}.
\newblock \emph{\bibinfo{journal}{\apjs}} \textbf{\bibinfo{volume}{224}},
  \bibinfo{pages}{33} (\bibinfo{year}{2016}).

\bibitem{Henault:2003}
\bibinfo{author}{{Henault}, F.} \emph{et~al.}
\newblock \bibinfo{title}{{MUSE: a second-generation integral-field
  spectrograph for the VLT}}.
\newblock In \bibinfo{editor}{{Iye}, M.} \& \bibinfo{editor}{{Moorwood},
  A.~F.~M.} (eds.) \emph{\bibinfo{booktitle}{Instrument Design and Performance
  for Optical/Infrared Ground-based Telescopes}}, vol. \bibinfo{volume}{4841}
  of \emph{\bibinfo{series}{\procspie}}, \bibinfo{pages}{1096--1107}
  (\bibinfo{year}{2003}).

\bibitem{Bacon:2012}
\bibinfo{author}{{Bacon}, R.} \emph{et~al.}
\newblock \bibinfo{title}{{News of the MUSE}}.
\newblock \emph{\bibinfo{journal}{The Messenger}}
  \textbf{\bibinfo{volume}{147}}, \bibinfo{pages}{4--6} (\bibinfo{year}{2012}).

\bibitem{Schmidt:2014}
\bibinfo{author}{{Schmidt}, K.~B.} \emph{et~al.}
\newblock \bibinfo{title}{{Through the Looking GLASS: HST Spectroscopy of Faint
  Galaxies Lensed by the Frontier Fields Cluster MACSJ0717.5+3745}}.
\newblock \emph{\bibinfo{journal}{\apjl}} \textbf{\bibinfo{volume}{782}},
  \bibinfo{pages}{L36} (\bibinfo{year}{2014}).

\bibitem{Treu:2015a}
\bibinfo{author}{{Treu}, T.} \emph{et~al.}
\newblock \bibinfo{title}{{The Grism Lens-Amplified Survey from Space (GLASS).
  I. Survey Overview and First Data Release}}.
\newblock \emph{\bibinfo{journal}{\apj}} \textbf{\bibinfo{volume}{812}},
  \bibinfo{pages}{114} (\bibinfo{year}{2015}).

\bibitem{Jullo:2007}
\bibinfo{author}{{Jullo}, E.} \emph{et~al.}
\newblock \bibinfo{title}{{A Bayesian approach to strong lensing modelling of
  galaxy clusters}}.
\newblock \emph{\bibinfo{journal}{New Journal of Physics}}
  \textbf{\bibinfo{volume}{9}}, \bibinfo{pages}{447} (\bibinfo{year}{2007}).

\bibitem{Kassiola:1993}
\bibinfo{author}{{Kassiola}, A.} \& \bibinfo{author}{{Kovner}, I.}
\newblock \bibinfo{title}{{Elliptic Mass Distributions versus Elliptic
  Potentials in Gravitational Lenses}}.
\newblock \emph{\bibinfo{journal}{\apj}} \textbf{\bibinfo{volume}{417}},
  \bibinfo{pages}{450} (\bibinfo{year}{1993}).

\bibitem{Limousin:2007}
\bibinfo{author}{{Limousin}, M.} \emph{et~al.}
\newblock \bibinfo{title}{{Combining Strong and Weak Gravitational Lensing in
  Abell 1689}}.
\newblock \emph{\bibinfo{journal}{\apj}} \textbf{\bibinfo{volume}{668}},
  \bibinfo{pages}{643--666} (\bibinfo{year}{2007}).

\bibitem{Kawamata:2016}
\bibinfo{author}{{Kawamata}, R.}, \bibinfo{author}{{Oguri}, M.},
  \bibinfo{author}{{Ishigaki}, M.}, \bibinfo{author}{{Shimasaku}, K.} \&
  \bibinfo{author}{{Ouchi}, M.}
\newblock \bibinfo{title}{{Precise Strong Lensing Mass Modeling of Four Hubble
  Frontier Field Clusters and a Sample of Magnified High-redshift Galaxies}}.
\newblock \emph{\bibinfo{journal}{\apj}} \textbf{\bibinfo{volume}{819}},
  \bibinfo{pages}{114} (\bibinfo{year}{2016}).

\bibitem{Oguri:2010b}
\bibinfo{author}{{Oguri}, M.}
\newblock \bibinfo{title}{{The Mass Distribution of SDSS J1004+4112
  Revisited}}.
\newblock \emph{\bibinfo{journal}{\pasj}} \textbf{\bibinfo{volume}{62}},
  \bibinfo{pages}{1017--} (\bibinfo{year}{2010}).

\bibitem{Suyu:2010b}
\bibinfo{author}{{Suyu}, S.~H.} \& \bibinfo{author}{{Halkola}, A.}
\newblock \bibinfo{title}{{The halos of satellite galaxies: the companion of
  the massive elliptical lens SL2S J08544-0121}}.
\newblock \emph{\bibinfo{journal}{\aap}} \textbf{\bibinfo{volume}{524}},
  \bibinfo{pages}{A94} (\bibinfo{year}{2010}).

\bibitem{Suyu:2012}
\bibinfo{author}{{Suyu}, S.~H.} \emph{et~al.}
\newblock \bibinfo{title}{{Disentangling Baryons and Dark Matter in the Spiral
  Gravitational Lens B1933+503}}.
\newblock \emph{\bibinfo{journal}{\apj}} \textbf{\bibinfo{volume}{750}},
  \bibinfo{pages}{10} (\bibinfo{year}{2012}).

\bibitem{Liesenborgs:2006}
\bibinfo{author}{{Liesenborgs}, J.}, \bibinfo{author}{{De Rijcke}, S.} \&
  \bibinfo{author}{{Dejonghe}, H.}
\newblock \bibinfo{title}{{A genetic algorithm for the non-parametric inversion
  of strong lensing systems}}.
\newblock \emph{\bibinfo{journal}{\mnras}} \textbf{\bibinfo{volume}{367}},
  \bibinfo{pages}{1209--1216} (\bibinfo{year}{2006}).

\bibitem{Liesenborgs:2007}
\bibinfo{author}{{Liesenborgs}, J.}, \bibinfo{author}{{de Rijcke}, S.},
  \bibinfo{author}{{Dejonghe}, H.} \& \bibinfo{author}{{Bekaert}, P.}
\newblock \bibinfo{title}{{Non-parametric inversion of gravitational lensing
  systems with few images using a multi-objective genetic algorithm}}.
\newblock \emph{\bibinfo{journal}{\mnras}} \textbf{\bibinfo{volume}{380}},
  \bibinfo{pages}{1729--1736} (\bibinfo{year}{2007}).

\bibitem{Mohammed:2014}
\bibinfo{author}{{Mohammed}, I.}, \bibinfo{author}{{Liesenborgs}, J.},
  \bibinfo{author}{{Saha}, P.} \& \bibinfo{author}{{Williams}, L.~L.~R.}
\newblock \bibinfo{title}{{Mass-galaxy offsets in Abell 3827, 2218 and 1689:
  intrinsic properties or line-of-sight substructures?}}
\newblock \emph{\bibinfo{journal}{\mnras}} \textbf{\bibinfo{volume}{439}},
  \bibinfo{pages}{2651--2661} (\bibinfo{year}{2014}).

\bibitem{Plummer:1911}
\bibinfo{author}{{Plummer}, H.~C.}
\newblock \bibinfo{title}{{On the problem of distribution in globular star
  clusters}}.
\newblock \emph{\bibinfo{journal}{\mnras}} \textbf{\bibinfo{volume}{71}},
  \bibinfo{pages}{460--470} (\bibinfo{year}{1911}).

\bibitem{Bradac:2005}
\bibinfo{author}{{Brada{\v c}}, M.}, \bibinfo{author}{{Schneider}, P.},
  \bibinfo{author}{{Lombardi}, M.} \& \bibinfo{author}{{Erben}, T.}
\newblock \bibinfo{title}{{Strong and weak lensing united}}.
\newblock \emph{\bibinfo{journal}{\aap}} \textbf{\bibinfo{volume}{437}},
  \bibinfo{pages}{39--48} (\bibinfo{year}{2005}).

\bibitem{Bradac:2009}
\bibinfo{author}{{Brada{\v c}}, M.} \emph{et~al.}
\newblock \bibinfo{title}{{Focusing Cosmic Telescopes: Exploring Redshift z
  \~{} 5-6 Galaxies with the Bullet Cluster 1E0657 - 56}}.
\newblock \emph{\bibinfo{journal}{\apj}} \textbf{\bibinfo{volume}{706}},
  \bibinfo{pages}{1201--1212} (\bibinfo{year}{2009}).

\bibitem{Sendra:2014}
\bibinfo{author}{{Sendra}, I.}, \bibinfo{author}{{Diego}, J.~M.},
  \bibinfo{author}{{Broadhurst}, T.} \& \bibinfo{author}{{Lazkoz}, R.}
\newblock \bibinfo{title}{{Enabling non-parametric strong lensing models to
  derive reliable cluster mass distributions - WSLAP+}}.
\newblock \emph{\bibinfo{journal}{\mnras}} \textbf{\bibinfo{volume}{437}},
  \bibinfo{pages}{2642--2651} (\bibinfo{year}{2014}).

\bibitem{Zitrin:2009a}
\bibinfo{author}{{Zitrin}, A.} \emph{et~al.}
\newblock \bibinfo{title}{{New multiply-lensed galaxies identified in ACS/NIC3
  observations of Cl0024+1654 using an improved mass model}}.
\newblock \emph{\bibinfo{journal}{\mnras}} \textbf{\bibinfo{volume}{396}},
  \bibinfo{pages}{1985--2002} (\bibinfo{year}{2009}).

\bibitem{Zitrin:2015}
\bibinfo{author}{{Zitrin}, A.} \emph{et~al.}
\newblock \bibinfo{title}{{Hubble Space Telescope Combined Strong and Weak
  Lensing Analysis of the CLASH Sample: Mass and Magnification Models and
  Systematic Uncertainties}}.
\newblock \emph{\bibinfo{journal}{\apj}} \textbf{\bibinfo{volume}{801}},
  \bibinfo{pages}{44} (\bibinfo{year}{2015}).

\bibitem{Mann:2012}
\bibinfo{author}{{Mann}, A.~W.} \& \bibinfo{author}{{Ebeling}, H.}
\newblock \bibinfo{title}{{X-ray-optical classification of cluster mergers and
  the evolution of the cluster merger fraction}}.
\newblock \emph{\bibinfo{journal}{\mnras}} \textbf{\bibinfo{volume}{420}},
  \bibinfo{pages}{2120--2138} (\bibinfo{year}{2012}).

\bibitem{Christensen:2012}
\bibinfo{author}{{Christensen}, L.} \emph{et~al.}
\newblock \bibinfo{title}{{The low-mass end of the fundamental relation for
  gravitationally lensed star-forming galaxies at 1 < z < 6}}.
\newblock \emph{\bibinfo{journal}{\mnras}} \textbf{\bibinfo{volume}{427}},
  \bibinfo{pages}{1953--1972} (\bibinfo{year}{2012}).

\bibitem{Henze:2015a}
\bibinfo{author}{{Henze}, M.} \emph{et~al.}
\newblock \bibinfo{title}{{A remarkable recurrent nova in M 31: The predicted
  2014 outburst in X-rays with Swift}}.
\newblock \emph{\bibinfo{journal}{\aap}} \textbf{\bibinfo{volume}{580}},
  \bibinfo{pages}{A46} (\bibinfo{year}{2015}).

\bibitem{Prialnik:1995}
\bibinfo{author}{{Prialnik}, D.} \& \bibinfo{author}{{Kovetz}, A.}
\newblock \bibinfo{title}{{An extended grid of multicycle nova evolution
  models}}.
\newblock \emph{\bibinfo{journal}{\apj}} \textbf{\bibinfo{volume}{445}},
  \bibinfo{pages}{789--810} (\bibinfo{year}{1995}).

\bibitem{Kato:2015}
\bibinfo{author}{{Kato}, M.}, \bibinfo{author}{{Saio}, H.} \&
  \bibinfo{author}{{Hachisu}, I.}
\newblock \bibinfo{title}{{Multi-wavelength Light Curve Model of the One-year
  Recurrence Period Nova M31N 2008-12A}}.
\newblock \emph{\bibinfo{journal}{\apj}} \textbf{\bibinfo{volume}{808}},
  \bibinfo{pages}{52} (\bibinfo{year}{2015}).

\bibitem{Chabrier:2003}
\bibinfo{author}{{Chabrier}, G.}
\newblock \bibinfo{title}{{Galactic Stellar and Substellar Initial Mass
  Function}}.
\newblock \emph{\bibinfo{journal}{\pasp}} \textbf{\bibinfo{volume}{115}},
  \bibinfo{pages}{763--795} (\bibinfo{year}{2003}).

\bibitem{Hogg:2002}
\bibinfo{author}{Hogg, D.~W.}, \bibinfo{author}{Baldry, I.~K.},
  \bibinfo{author}{Blanton, M.~R.} \& \bibinfo{author}{Eisenstein, D.~J.}
\newblock \bibinfo{title}{The k correction}.
\newblock \emph{\bibinfo{journal}{arXiv:astro-ph/0210394}}
  (\bibinfo{year}{2002}).

\bibitem{Graur:2014}
\bibinfo{author}{{Graur}, O.} \emph{et~al.}
\newblock \bibinfo{title}{{Type-Ia Supernova Rates to Redshift 2.4 from CLASH:
  The Cluster Lensing And Supernova Survey with Hubble}}.
\newblock \emph{\bibinfo{journal}{\apj}} \textbf{\bibinfo{volume}{783}},
  \bibinfo{pages}{28} (\bibinfo{year}{2014}).

\bibitem{Rodney:2014}
\bibinfo{author}{{Rodney}, S.~A.} \emph{et~al.}
\newblock \bibinfo{title}{{Type Ia Supernova Rate Measurements to Redshift 2.5
  from CANDELS: Searching for Prompt Explosions in the Early Universe}}.
\newblock \emph{\bibinfo{journal}{\aj}} \textbf{\bibinfo{volume}{148}},
  \bibinfo{pages}{13} (\bibinfo{year}{2014}).


\end{thebibliography}

\begin{thebibliography}{100}
\makeatletter
\addtocounter{\@listctr}{86}
\makeatother

\expandafter\ifx\csname url\endcsname\relax
  \def\url#1{\texttt{#1}}\fi
\expandafter\ifx\csname urlprefix\endcsname\relax\def\urlprefix{URL }\fi
\providecommand{\bibinfo}[2]{#2}
\providecommand{\eprint}[2][]{\url{#2}}


\bibitem{Sersic:1963}
\bibinfo{author}{{S{\'e}rsic}, J.~L.}
\newblock \bibinfo{title}{{Influence of the atmospheric and instrumental
  dispersion on the brightness distribution in a galaxy}}.
\newblock \emph{\bibinfo{journal}{Boletin de la Asociacion Argentina de
  Astronomia La Plata Argentina}} \textbf{\bibinfo{volume}{6}},
  \bibinfo{pages}{41} (\bibinfo{year}{1963}).

\bibitem{Akaike:1974}
\bibinfo{author}{Akaike, H.}
\newblock \bibinfo{title}{A new look at the statistical model identification}.
\newblock \emph{\bibinfo{journal}{Automatic Control, IEEE Transactions on}}
  \textbf{\bibinfo{volume}{19}}, \bibinfo{pages}{716--723}
  (\bibinfo{year}{1974}).

\bibitem{Hemmati:2014}
\bibinfo{author}{{Hemmati}, S.} \emph{et~al.}
\newblock \bibinfo{title}{{Kiloparsec-scale Properties of Emission-line
  Galaxies}}.
\newblock \emph{\bibinfo{journal}{\apj}} \textbf{\bibinfo{volume}{797}},
  \bibinfo{pages}{108} (\bibinfo{year}{2014}).

\bibitem{Wambsganss:2001}
\bibinfo{author}{{Wambsganss}, J.}
\newblock \bibinfo{title}{{Quasar Microlensing}}.
\newblock In \bibinfo{editor}{{Brainerd}, T.~G.} \&
  \bibinfo{editor}{{Kochanek}, C.~S.} (eds.)
  \emph{\bibinfo{booktitle}{Gravitational Lensing: Recent Progress and Future
  Go}}, vol. \bibinfo{volume}{237} of \emph{\bibinfo{series}{Astronomical
  Society of the Pacific Conference Series}}, \bibinfo{pages}{185}
  (\bibinfo{year}{2001}).

\bibitem{Kochanek:2004}
\bibinfo{author}{{Kochanek}, C.~S.}
\newblock \bibinfo{title}{{Quantitative Interpretation of Quasar Microlensing
  Light Curves}}.
\newblock \emph{\bibinfo{journal}{\apj}} \textbf{\bibinfo{volume}{605}},
  \bibinfo{pages}{58--77} (\bibinfo{year}{2004}).

\bibitem{Paczynski:1986}
\bibinfo{author}{{Paczynski}, B.}
\newblock \bibinfo{title}{{Gravitational microlensing by the galactic halo}}.
\newblock \emph{\bibinfo{journal}{\apj}} \textbf{\bibinfo{volume}{304}},
  \bibinfo{pages}{1--5} (\bibinfo{year}{1986}).

\bibitem{Alcock:1993}
\bibinfo{author}{{Alcock}, C.} \emph{et~al.}
\newblock \bibinfo{title}{{Possible gravitational microlensing of a star in the
  Large Magellanic Cloud}}.
\newblock \emph{\bibinfo{journal}{\nat}} \textbf{\bibinfo{volume}{365}},
  \bibinfo{pages}{621--623} (\bibinfo{year}{1993}).

\bibitem{Aubourg:1993}
\bibinfo{author}{{Aubourg}, E.} \emph{et~al.}
\newblock \bibinfo{title}{{Evidence for gravitational microlensing by dark
  objects in the Galactic halo}}.
\newblock \emph{\bibinfo{journal}{\nat}} \textbf{\bibinfo{volume}{365}},
  \bibinfo{pages}{623--625} (\bibinfo{year}{1993}).

\bibitem{Udalski:1993}
\bibinfo{author}{{Udalski}, A.} \emph{et~al.}
\newblock \bibinfo{title}{{The optical gravitational lensing experiment.
  Discovery of the first candidate microlensing event in the direction of the
  Galactic Bulge}}.
\newblock \emph{\bibinfo{journal}{ACTAA}} \textbf{\bibinfo{volume}{43}},
  \bibinfo{pages}{289--294} (\bibinfo{year}{1993}).

\bibitem{Chang:1979}
\bibinfo{author}{{Chang}, K.} \& \bibinfo{author}{{Refsdal}, S.}
\newblock \bibinfo{title}{{Flux variations of QSO 0957+561 A, B and image
  splitting by stars near the light path}}.
\newblock \emph{\bibinfo{journal}{\nat}} \textbf{\bibinfo{volume}{282}},
  \bibinfo{pages}{561--564} (\bibinfo{year}{1979}).

\bibitem{Chang:1984}
\bibinfo{author}{{Chang}, K.} \& \bibinfo{author}{{Refsdal}, S.}
\newblock \bibinfo{title}{{Star disturbances in gravitational lens galaxies}}.
\newblock \emph{\bibinfo{journal}{\aap}} \textbf{\bibinfo{volume}{132}},
  \bibinfo{pages}{168--178} (\bibinfo{year}{1984}).

\bibitem{Darnley:2014}
\bibinfo{author}{{Darnley}, M.~J.} \emph{et~al.}
\newblock \bibinfo{title}{{A remarkable recurrent nova in M 31: The optical
  observations}}.
\newblock \emph{\bibinfo{journal}{\aap}} \textbf{\bibinfo{volume}{563}},
  \bibinfo{pages}{L9} (\bibinfo{year}{2014}).

\bibitem{Planck:2016}
\bibinfo{author}{{Planck Collaboration}} \emph{et~al.}
\newblock \bibinfo{title}{{Planck 2015 results. XIII. Cosmological
  parameters}}.
\newblock \emph{\bibinfo{journal}{\aap}} \textbf{\bibinfo{volume}{594}},
  \bibinfo{pages}{A13} (\bibinfo{year}{2016}).


\end{thebibliography}


\clearpage
\begin{supplementary}

\subsection{Lens Model Variations.}\label{sec:LensModelVariations}

Supplementary Figure~\ref{fig:LensModelContours} presents probability
distributions for the three magnifications and two time delay values
of interest.  These distributions were derived by combining the Monte
Carlo chains from the CATS, GLAFIC, GLEE, and ZLTM models, and
individual runs of the GRALE model, which uses a different random seed
for each run.  We applied a weight to each model to account for the
different number of model iterations used by each modeling team. All
five of these models agree that host image 11.3 is the leading image,
appearing some 3--7 years before the other two images.  The models do
not agree on the arrival sequence of images 11.1 and 11.2: some have
the NW image 11.2 as a leading image, and others have it as a trailing
image.  However, the models do consistently predict that the
separation in time between those two images should be roughly in the
range of 1 to 60 days.

Because of the proximity of the critical curves in all models, the
predicted time delays and magnification factors are significantly
different if calculated at the model-predicted positions instead of
the observed positions.  For example, in the GLEE model series (GLEE
and GLEE-var) when switching from the observed to model-predicted
positions the arrival order of the NW and SE images flips, the
expected time delay drops from tens of days to $<$1 day, and the
magnifications change by 30-60\%.  However, the expected
magnifications and time delays between the events still fall within
the broad ranges summarized in Table~\ref{tab:LensModelPredictions}
and shown in Supplementary Figure~\ref{fig:LensModelContours}.  Regardless of
whether the model predictions are extracted at the observed or
predicted positions of the \spock events, none of the lens models can
accommodate the observed 234-day time difference as purely a
gravitational lensing time delay.

We used variations of several lens models to investigate how the
lensing critical curves shift under a range of alternative assumptions
or input constraints.  These variations highlight the range of
systematic effects that might impact the model predictions for the
\spock magnifications, time delays and proximity to the critical
curves.  Figure~\ref{fig:SpockCriticalCurves} shows the critical
curves for a source at $z=1$ (the redshift of the \spock host galaxy)
predicted by our seven baseline models, plus the four variations
described below.  Within a given model, variations that move a
critical curve closer to the position of \spockone\ would drive the
magnification of that event much higher (toward $\mu_{\rm
  NW}\approx200$).  This generally also has the effect of moving the
critical curve farther from \spocktwo, which would necessarily drive
its magnification downward (toward $\mu_{\rm SE}\approx10$).

The baseline CATS model reported in
Table~\ref{tab:LensModelPredictions} corresponds to the CATSv4.1 model
published on the STScI Frontier Fields lens model repository
(\url{https://archive.stsci.edu/pub/hlsp/frontier/macs0416/models/cats/v4.1/}).
That model uses 178 cluster member galaxies, including a galaxy $<5''$
south of the \spock host galaxy, which creates a local critical curve
that intersects the \spocktwo location.  Our CATS-var model is an
earlier iteration of the model, published on the STScI repository as
CATSv4
(\url{https://archive.stsci.edu/pub/hlsp/frontier/macs0416/models/cats/v4/}),
and includes only 98 galaxies identified as cluster members.  In this
variation the nearby cluster member galaxy is not included, so the
\spocktwo event is not intersected by a critical curve. However, the
\spockone event is approximately coincident with the primary critical
curve of the \macs0416 cluster.  When the critical curve is close to
either \spock location, the magnifications predicted by the CATS model
are driven up to $\mu>100$.  However, the time delays remain small, on
the order of tens of days, and incompatible with the observed 234-day
gap.

The WSLAP-var model evaluates whether the cluster redshift
significantly impacts the positioning of the critical curve. In this
merging cluster, the northern brightest cluster galaxy (BCG) has a
slighter higher redshift than the southern BCG. The mean redshift of
the cluster is not precisely determined, since it is likely to be
aligned somewhat along the line of sight.  For the WSLAP-var model we
shift the assumed cluster redshift $z=0.4$ from the default $z=0.396$
(used in all the baseline models).  The shift in the critical curve is
noticeable, but not substantial, insofar as this change does not drive
the critical curve to intersect either or both of the \spock
locations.

The GLEE-var model is a multi-plane lens model (Chiriv{\`i} et al.,~in
prep.) that incorporates 13 galaxies with spectroscopic redshifts that
place them either in the foreground or background of the \macs0416
cluster.  Supplementary Figure~\ref{fig:LineOfSightLenses} marks these
13 galaxies and highlights two of them that appear in the foreground
of the \spock host galaxy and are close to the lines of sight to the
\spock transients. Both the foreground $z=0.0557$ galaxy and the
reconstructed position of the $z=0.9397$ galaxy have a projected
separation of $<$4\arcsec from the \spocktwo transient position.
Including these galaxies in the GLEE lensing model changes the
absolute value of the magnifications at the location of HFF14Spo-NW
(HFF14Spo-SE) to $\sim70$ ($\sim250$) and the time delay between the
two locations to $\sim50$ days.  The line-of-sight galaxies also
result in a shift of the position of the critical curve---as can be
seen by comparing the GLEE and GLEE-var models in
Figure~\ref{fig:SpockCriticalCurves}. Nonetheless, the predicted time
delays are still incompatible with the observed gap of 234 days
between events.

The GLAFIC-var model examines whether it is plausible for a critical
curve to intersect both \spock locations---contrary to the baseline
assumption of a single critical curve subtending the \spock host
galaxy roughly midway between the two positions.  This model includes
a customized constraint, requiring that the magnification factors at
the \spock positons are $>1000$.  To achieve this, we independently
adjusted the mass scaling for the two nearest cluster member galaxies,
which are located just northeast and south of the \spock host galaxy
arc.  The mass of the northeast member galaxy was increased by
$\sim$30\% and that of the southern one by $\sim$60\%.  As a simple
check of the predicted morphology of the host galaxy, we placed a
source with a simple Sersic profile\cite{Sersic:1963} on the source
plane. The lensed image of that artificial source is an unbroken
elongated arc, reproducing the host galaxy image morphology reasonably
well.

For this modification of the GLAFIC lens model to be justified in a
statistical sense, the revised model should still accurately reproduce
the observed strong-lensing constraints across the entire cluster.
The $\chi^2$ statistic for the baseline GLAFIC model is 240, with 196
degrees of freedom ($\chi^2_\nu=1.2$), and yields an Akaike
information criterion (AIC)\citep{Akaike:1974} of 676.  For the
GLAFIC-var model that forces multiple critical curves to intersect the
\spock locations, we get $\chi^2$=331 for 192 degrees of freedom
($\chi^2_\nu=1.7$) and AIC=769.  This suggests that the multiple
critical curve GLAFIC-var model is strongly disfavored by the
{\it positional} strong-lensing constraints that are used for both models.
However, we note that neither model incorporates the temporal
constraints of the observed time delay. 

A second variation of the CATS model (CATS-var2) was also used to test
the plausibility of multiple critical curves intersecting the \spock
locations.  As in the GLAFIC-var case, this model requires that
critical curves pass very near the \spock positions.  The model can
accommodate that constraint, insofar as the root mean square (RMS)
error of the best-fit model is similar to that of the CATS and
CATS-var models.  However, in this CATS-var2 model the \spock host
galaxy is predicted to be multiply-imaged 5 times. The \HST images do
not exhibit any breaks or substructure in the arc that would be
generally expected in such a situation.  

Moreover, this CATS-var2 model has strong implications for a separate
background galaxy in the vicinity of image 11.3 (system 14 in
\citeref{Caminha:2017}).  This galaxy is strongly lensed by a pair of
spectroscopically confirmed cluster member galaxies\citep{Caminha:2017}. Comparing the observed positions of the multiple
images of System 14 against the CATS-var2 model-predicted positions,
we find that this System contributes significantly to the global RMS
error for the model---indicating that the CATS-var2 model can not
accurately reproduce the multiple images of System 14.  Conversely,
when this system is removed as a model constraint, the RMS error
decreases, and the CATS-var2 model can more successfully pass the
critical line through the two \spock locations.  A possible
interpretation of this is that the established strong lensing
constraints (especially System 14) are incompatible with the
requirement that multiple critical curves must intersect the two
\spock locations.

\subsection{Host Galaxy}\label{sec:HostGalaxy}

To examine whether the two transients originated from the same
physical location in the source plane, we looked for differences in
the properties of the \spock host galaxy at the location of each
event.  We first used the technique of ``pixel-by-pixel'' SED
fitting\cite{Hemmati:2014} to determine rest-frame colors and stellar
properties in a single resolution element of the \HST imaging data.
For this purpose we used the deepest possible stacks of \HST images,
comprising all available data except those images where the transient
events were present.  The resulting maps of stellar population
properties are shown in Supplementary Figure~\ref{fig:HostProperties}.
Supplementary Table~\ref{tab:HostProperties} reports measurements of
the three derived stellar population properties (color, mass, age)
from host images 11.1, 11.2 and 11.3.  In 11.1 and 11.2 these
measurements were extracted from the central pixel at the location of
each of the two \spock events.  The lensing magnification here ranges
from $\mu=10$ to 200, corresponding to a size on the source plane
between 6 and 600 pc$^2$.  For host image 11.3 we report the stellar
population properties derived from the pixel at the center of the
galaxy, because the lens models do not have sufficient precision to
map the \spock locations to specific positions in image 11.3.  With a
magnification of $\sim$3 to 5, this extraction region covers roughly
2000 to 6000 pc$^2$.

The reported uncertainties for these derived stellar properties in
Table~\ref{tab:HostProperties} reflect only the measurement errors
from the SED fitting, and do not attempt to quantify potential
systematic biases.  Such biases could arise, for example, from color
differences in the background light, which is dominated by the cluster
galaxies and varies significantly across the \MACS0416 field.  Such a
bias might shift the absolute values of the parameter scales for any
given host image (e.g., making the galaxy as a whole appear bluer,
more massive, and younger). However, the gradients across any single
host image are unlikely to be driven primarily by such systematics.

Supplementary Figure~\ref{fig:HostProperties} and Supplementary
Table~\ref{tab:HostProperties} show that the measured values of the
color, stellar mass, and age at the two \spock locations are mutually
consistent. Thus, it is plausible to assume that the two positions map
back to the same physical location at the source plane.  Comparing
those two locations to the center of the galaxy as defined in image
11.3, we see only a mild tension in the rest frame $U-V$ color. This
comparison therefore cannot quantitatively rule out the possibility
that the two transient events are located at the center of the
galaxy. However, the maps shown in Supplementary Figure~\ref{fig:HostProperties} do
show a gradient in both $U-V$ color and stellar age. For both images
11.1 and 11.2 the bluest and youngest stars ($U-V\approx0.3$,
$\tau\approx280$ Myr) are localized in knots near the extreme ends of
each image, well separated from either of the \spock transient events.
In the less distorted host image 11.3 the bluer and younger stars are
concentrated near the center. Taken together, these color and age
gradients suggest that the two transients are not coincident with the
center of their host galaxy.

\begin{table}
\caption{Measurements of the \forbidden{O}{ii} $\lambda\lambda$3626,3629 lines from \spock host galaxy images 11.1 and 11.2}
\resizebox{\textwidth}{!}{%
\begin{tabular}{rllc ccc ccc c}
\tableline
\tableline
  \multicolumn{1}{c}{Aperture} & \multicolumn{1}{c}{R.A.} & \multicolumn{1}{c}{Dec.} & \multicolumn{1}{c}{distance to} &
  \multicolumn{3}{c}{\forbidden{O}{ii} $\lambda$3726} & \multicolumn{3}{c}{\forbidden{O}{ii} $\lambda$3729} & \multicolumn{1}{c}{Line}\\ 
  \multicolumn{1}{c}{ID} & \multicolumn{1}{c}{J2000} & \multicolumn{1}{c}{J2000} & \multicolumn{1}{c}{\spocktwo} &
  \multicolumn{1}{c}{Flux} & \multicolumn{1}{c}{$\lambda_{\rm center}$} & \multicolumn{1}{c}{FWHM} &
  \multicolumn{1}{c}{Flux} & \multicolumn{1}{c}{$\lambda_{\rm center}$} & \multicolumn{1}{c}{FWHM} & \multicolumn{1}{c}{Ratio}\\
  & \multicolumn{1}{c}{(degrees)} & \multicolumn{1}{c}{(degrees)} & \multicolumn{1}{c}{(Arcsec)} &
  \multicolumn{1}{c}{(erg\,s$^{-1}$\,cm$^{-2}$)} & \multicolumn{1}{c}{(\AA)} & \multicolumn{1}{c}{(\AA)} &
  \multicolumn{1}{c}{(erg\,s$^{-1}$\,cm$^{-2}$)} & \multicolumn{1}{c}{(\AA)} & \multicolumn{1}{c}{(\AA)}\\[0.5em]
\tableline\\
    1  & 64.039371 &  -24.070450 &  -1.54 & 2.19e-18 &   7472.37 &  4.00 &    3.57e-18 &   7478.17 & 4.00 &      1.63\\      
    2  & 64.039218 &  -24.070345 &  -0.88 & 4.73e-18 &   7472.16 &  4.00 &    5.30e-18 &   7478.12 & 3.40 &      1.12\\
    3  & 64.039078 &  -24.070264 &  -0.30 & 5.05e-18 &   7472.29 &  4.00 &    6.10e-18 &   7478.27 & 3.73 &      1.21\\
    4  & 64.038921 &  -24.070163 &   0.39 & 4.22e-18 &   7472.19 &  4.00 &    5.74e-18 &   7478.08 & 3.59 &      1.36\\
    5  & 64.038785 &  -24.070078 &   0.97 & 3.86e-18 &   7472.25 &  4.00 &    6.56e-18 &   7478.19 & 4.00 &      1.70\\
    6  & 64.038637 &  -24.069958 &   1.65 & 4.80e-18 &   7472.51 &  4.00 &    5.42e-18 &   7478.07 & 2.69 &      1.13\\
    7  & 64.038501 &  -24.069865 &   2.24 & 4.60e-18 &   7472.57 &  3.43 &    5.74e-18 &   7478.17 & 3.20 &      1.25\\
    8  & 64.038352 &  -24.069752 &   2.92 & 4.70e-18 &   7472.54 &  3.54 &    6.22e-18 &   7478.16 & 2.95 &      1.32\\
    9  & 64.038229 &  -24.069648 &   3.50 & 3.26e-18 &   7472.83 &  2.80 &    5.79e-18 &   7478.16 & 2.84 &      1.77\\
   10  & 64.038076 &  -24.069532 &   4.19 & 2.44e-18 &   7473.01 &  2.57 &    3.22e-18 &   7478.10 & 2.73 &      1.32\\
   NW  & 64.038565 &  -24.069939 &   1.90 & 4.30e-18 &   7472.55 &  3.13 &    5.49e-18 &   7478.01 & 2.89 &      1.28\\
   SE  & 64.038998 &  -24.070241 &   0.00 & 4.37e-18 &   7472.46 &  4.00 &    6.10e-18 &   7478.22 & 3.79 &      1.40\\
\tableline
\end{tabular}}
\label{tab:MuseLineFits}
\end{table}

In addition to the \HST imaging data, we also have spatially resolved
spectroscopy from the MUSE integral field data.  The only significant
spectral line feature for the \spock host is the \forbidden{O}{ii}
$\lambda\lambda$ 3726, 3729 doublet, observed at 7474 and 7478 \AA.
To examine this feature in detail, one-dimensional spectra were
extracted from the three-dimensional MUSE data cube at a series of
locations along the \spock host-galaxy arc.

Supplementary Figure~\ref{fig:MUSEOIISequence} depicts the apertures used for these
extractions, shows the observed \forbidden{O}{ii} lines at the
\spock-NW and SE positions, and compares the \forbidden{O}{ii} line
profiles to other positions along the length of the host-galaxy arc.
At each position the lines were extracted using apertures with a
radius of $0.6''$, so adjacent extractions are not independent,
although the two extractions centered on the \spockone and -SE
positions have no overlap.  

A difference in the shape of the \forbidden{O}{ii} lines or the
doublet line ratio could provide evidence for a different environment
at the two \spock locaions, which would suggest that the two events
emerged from independent sources.  For a visual test for spectral
deviations, we first constructed a mean spectrum by averaging the 1-D
spectra from five non-overlapping apertures (apertures 1, 3, 5, 7,
9). To account for differences in magnification and host-galaxy
intensity across the arc, each input spectrum was normalized at the
wavelength 7477.7 \AA, which corresponds to the center of the
$\lambda$3729 component of the \forbidden{O}{ii} emission line.  This
mean spectrum was then subtracted from the 1-D spectrum of each
aperture, producing a set of ``residual spectra,'' shown in
Supplementary Figure~\ref{fig:MUSEOIISequence} in the lower-left panel.  These
spectra show no indication of a systematic trend in the wavelength
position, shape or line ratio across the arc.  Similarly, a comparison
of the spectra from the \spock-NW and SE locations (right panels of
Supplementary Figure~\ref{fig:MUSEOIISequence}) reveals no significant difference in
the \forbidden{O}{ii} line shapes.

This qualitative comparison is borne out by a more quantitative
assessment, reported in Table~\ref{tab:MuseLineFits}. We fit a
Gaussian profile to each component of the \forbidden{O}{ii} doublet,
separately in each extracted 1-D spectrum. From these fits we measured
the integrated line flux, observed wavelength of line center
($\lambda_{\rm center}$), full width at half maximum intensity (FWHM), and the
intensity ratio of the two components of the doublet.  These
quantities---all reported in Table~\ref{tab:MuseLineFits}---do not
exhibit any discernible gradient across the host galaxy.  Thus, the
\forbidden{O}{ii} measurements from MUSE cannot be used to distinguish
either \spock location from the other, or to definitively answer
whether either position is coincident with the center of the host
galaxy (as would be required, for example, if these transients were
from an AGN).  We conclude that it is entirely plausible but not
certain that the two \spock events arose from the same physical
location in the host galaxy.

\subsection{LBV Light-Curve Comparison.}
\label{sec:LBVlightcurves}

Supplementary Figure~\ref{fig:LBVLightCurveComparison} presents a direct
comparison of the observed \spock light curves against the light
curves of the two LBVs that have well-studied rapid eruptions: SN
2009ip and NGC3432-LBV1. The brief outbursts of these LBVs have been
less finely sampled than the two \spock events, but the available data
show a wide variety of rise and decline times, even for a single
object over a relatively narrow time window of a few months.

\subsection{Expected Timescale for Microlensing Events.}\label{sec:Microlensing}

A commonly observed example of microlensing-induced transient effects
is when a bright background source (a quasar) is magnified by a
galaxy-scale lens\citep{Wambsganss:2001, Kochanek:2004}.  In this
optically thick microlensing regime, the lensing potential along the
line of sight to the quasar is composed of many stellar-mass objects.
Each compact object along the line of sight generates a separate
critical lensing curve, resulting in a complex web of overlapping
critical curves. As all of these lensing stars are in motion relative
to the background source, the web of caustics will shift across the
source position, leading to a stochastic variability on timescales of
months to years.  This scenario is inconsistent with the observed
data, as the two \spock events were far too short in duration and did
not exhibit the repeated ``flickering'' variation that would be
expected from optically thick microlensing.

For the cluster-scale lens relevant in the case of \spock, we should
expect to be in the optically thin microlensing regime.  This
situation is similar to the ``local'' microlensing light curves
observed when stars within our Galaxy or neighboring dwarf galaxies
pass behind a massive compact halo object\citep{Paczynski:1986,
  Alcock:1993, Aubourg:1993, Udalski:1993}.  In this case, an isolated
microlensing event can occur if there is a background star (i.e., in
the \spock host galaxy) that is the dominant source of luminosity in
its environment. In practice this means that the source must be a very
bright O or B star with mass of order 10 \Msun.  Depending on its age,
the size of such a star would range from a few to a few dozen times
the size of the Sun.  The net relative transverse velocity would be on
the order of a few 100 km s$^{-1}$, which is comparable to the orbital
velocity of stars within a galaxy or galaxies within a cluster.

In the case of a smooth cluster potential, the timescale $\tau$ for
the light curve of such a caustic crossing event is dictated by the
radius of the source, $R$, and the net transverse velocity, $v$, of
the source across the caustic\citep{Chang:1979,Chang:1984,MiraldaEscude:1991} as

\begin{equation}
  \tau = \frac{6 R}{5\,\Rsun}\frac{300~{\rm km~ s}^{-1}}{v}~\rm{hr.}
\label{eqn:caustic_crossing_time}
\end{equation}

\noindent Thus, for reasonable assumptions about the star's radius and
velocity, the timescale $\tau$ is on the order of hours to days, which is well
matched to the observed rise and decline timescales of the \spock
events.

\begin{figure*}[tbp]
  \begin{center}
    \includegraphics[width=\textwidth]{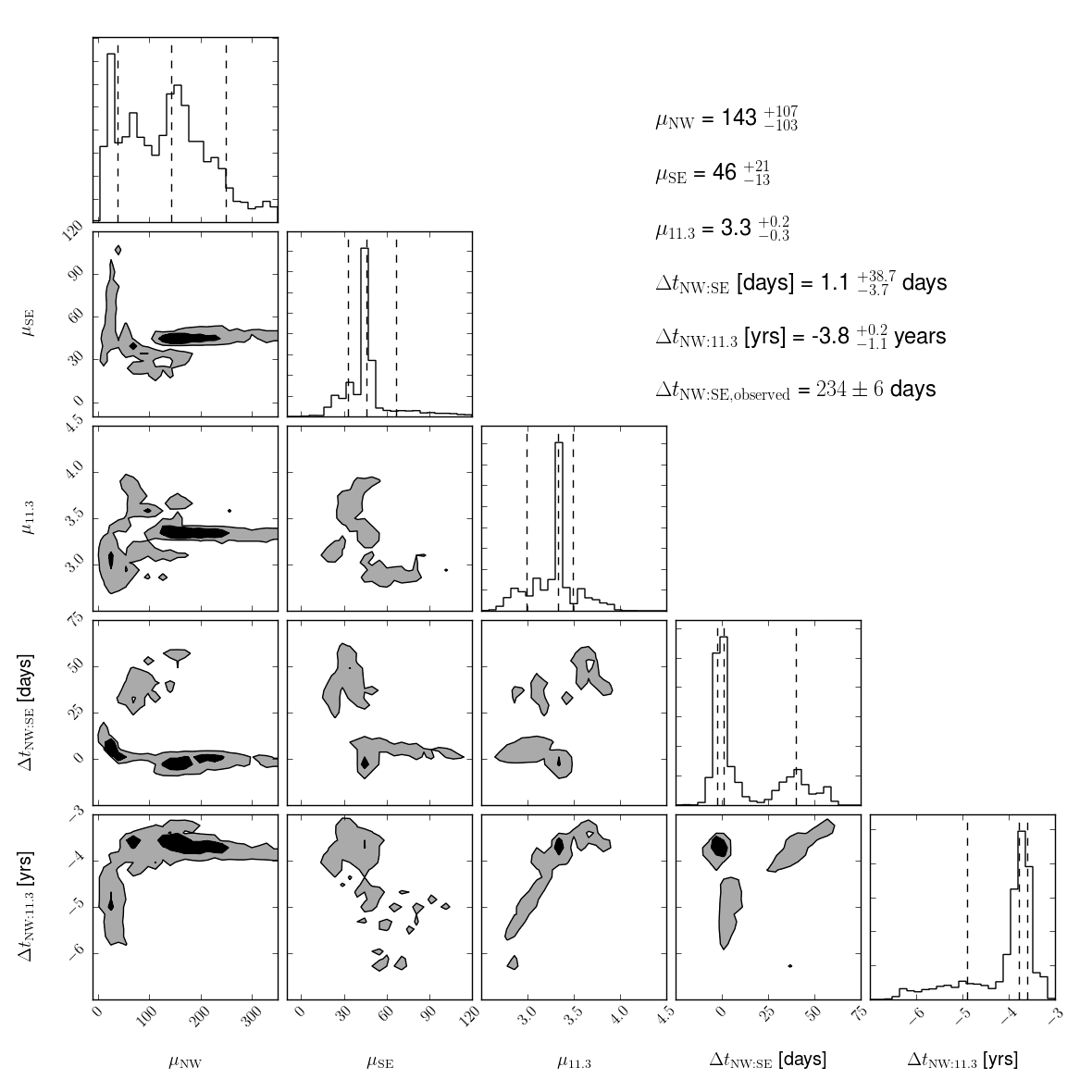}
    \caption{ \protect\input{composite_lens_model_contours_caption.tex}}
  \end{center}
\end{figure*}

\begin{figure*}[tbp]
  \begin{center}
    \includegraphics[width=\textwidth]{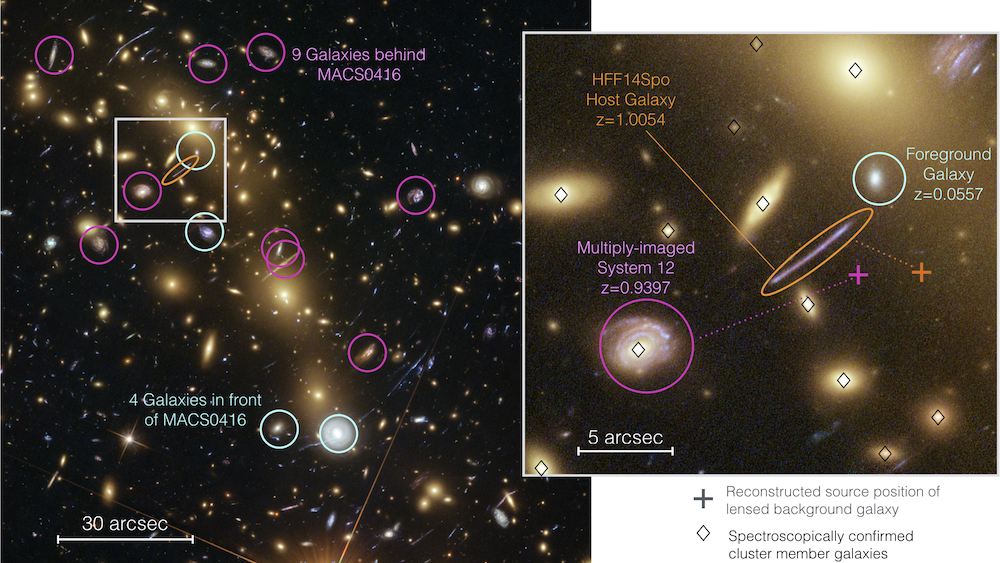}
    \caption{\protect\input{macs0416_lineofsight_lensing_caption.tex}}
  \end{center}
\end{figure*}

\begin{figure*}[tbp]
  \begin{center}
    \includegraphics[width=0.5\textwidth]{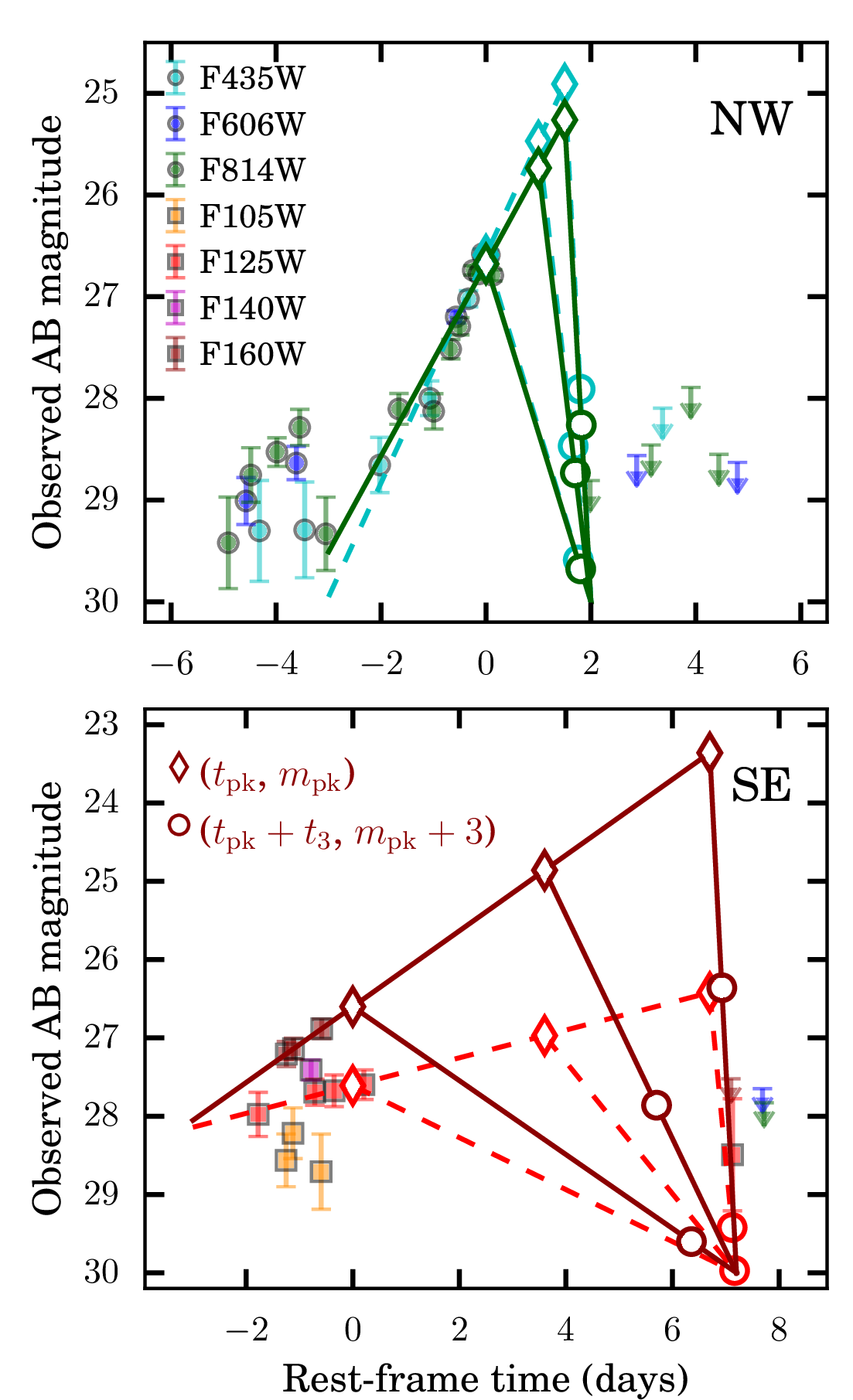}
    \caption{\protect\input{light_curve_linear_fits_caption.tex}}
  \end{center}
\end{figure*}

\begin{figure*}[tbp]
  \begin{center}
    \includegraphics[width=\textwidth]{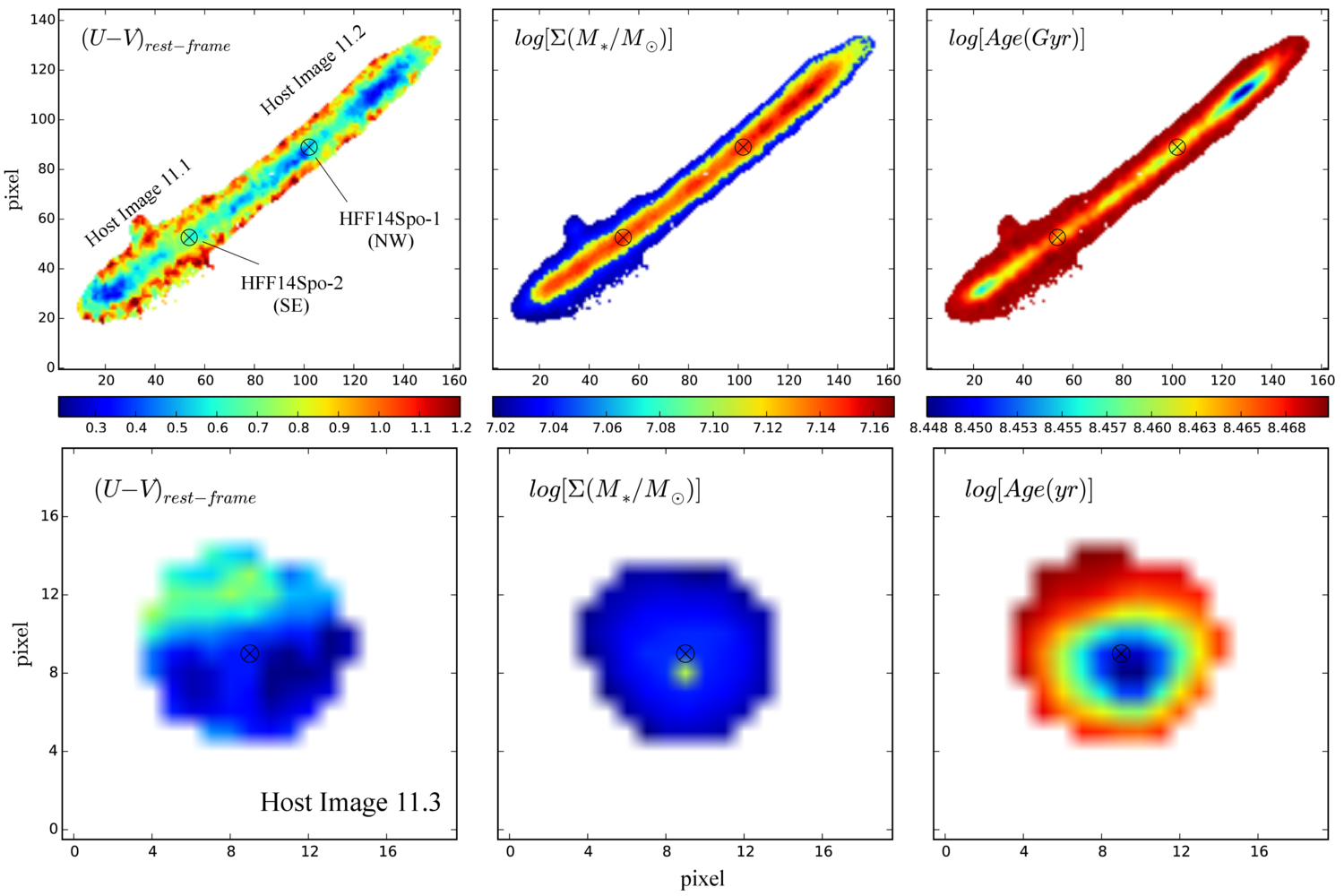}
    \caption{\protect\input{spock_hostgalaxy_properties_caption.tex}}
  \end{center}
\end{figure*}

\begin{figure*}[tbp]
  \begin{center}
    \includegraphics[width=\textwidth]{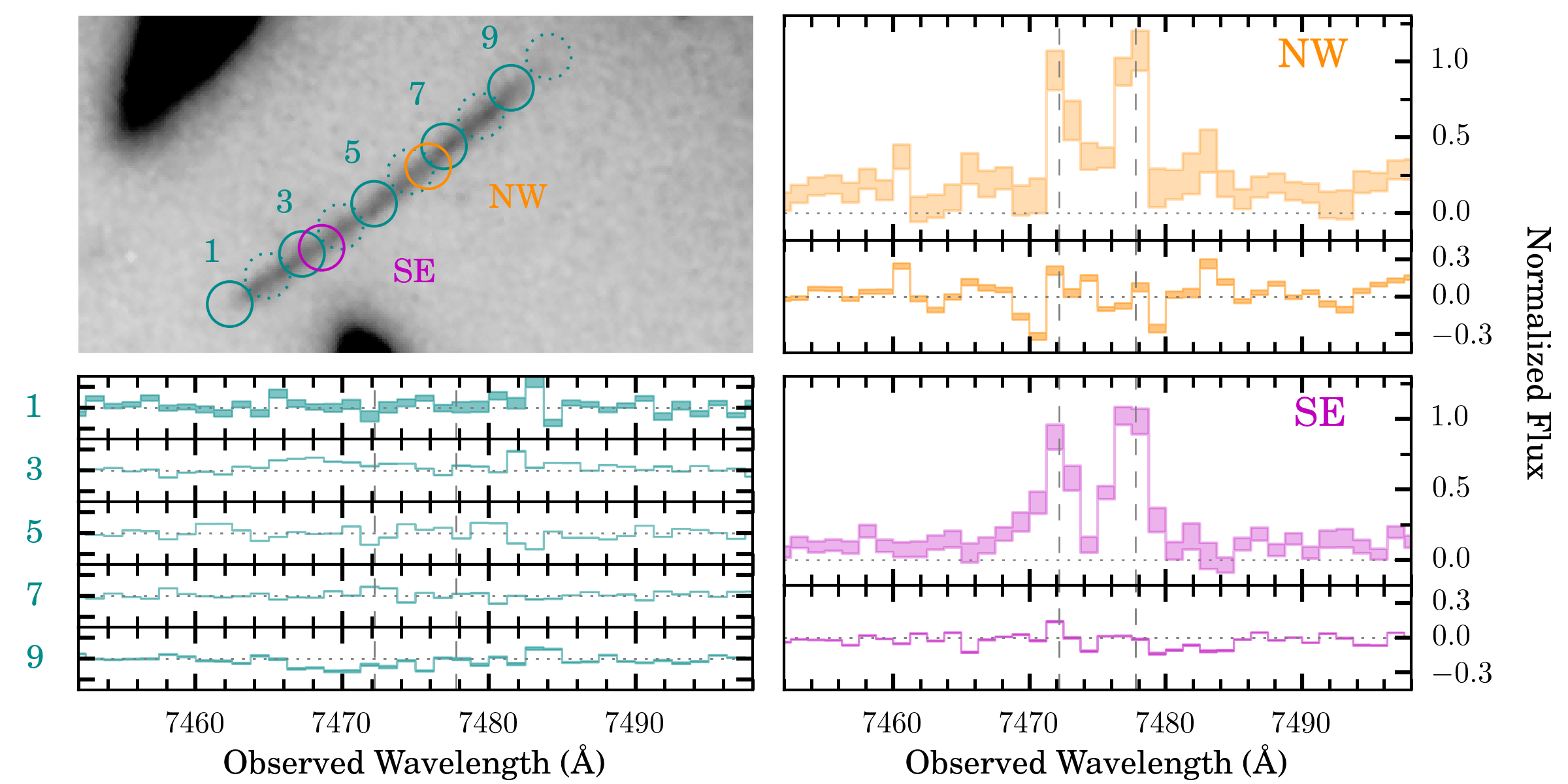}
    \caption{\protect\input{muse_oii_sequence_caption.tex}}
  \end{center}
\end{figure*}

\begin{figure*}[tbp]
  \begin{center}
    \includegraphics[width=\textwidth]{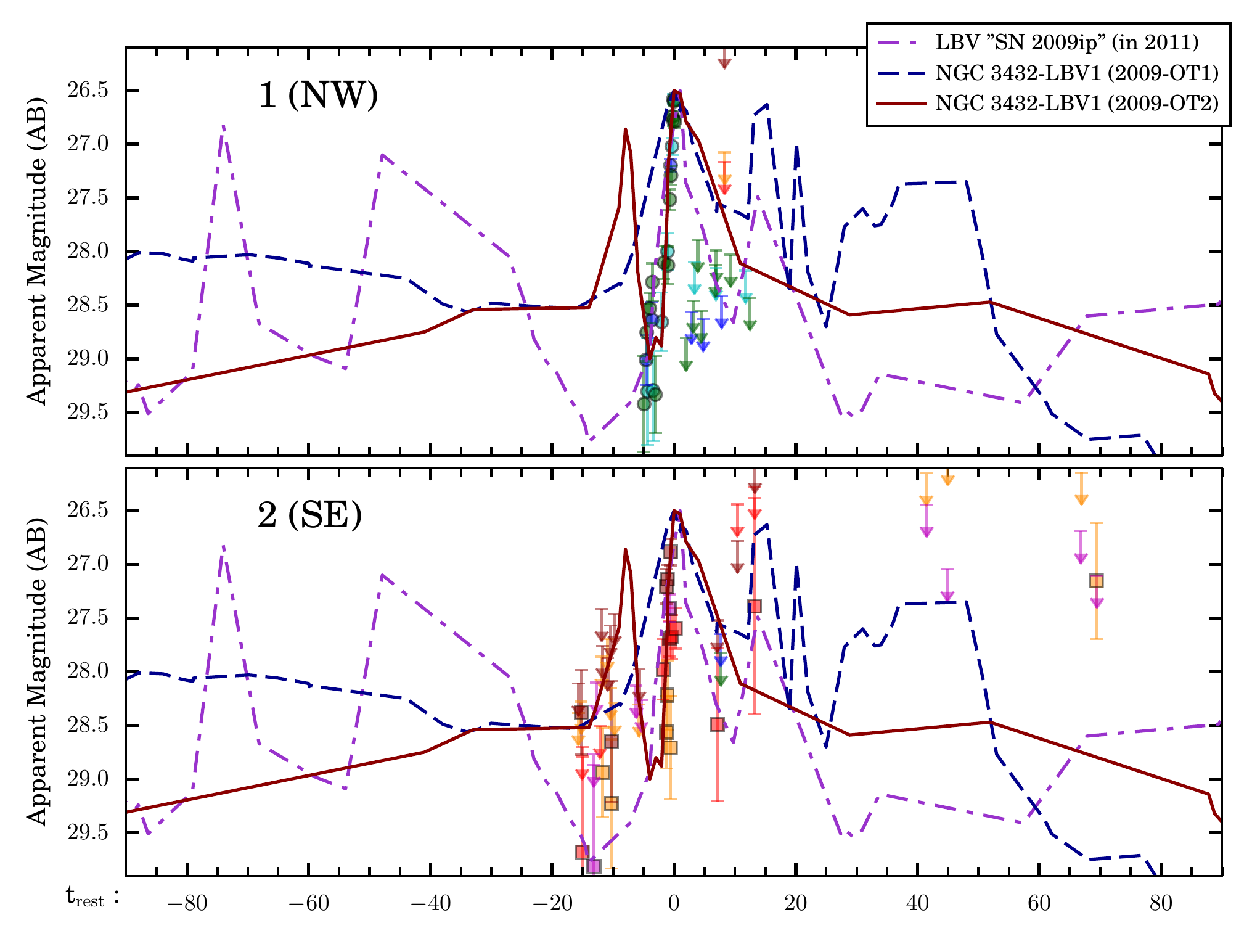}
    \caption{\protect\input{lbv_lightcurve_comparison_caption.tex}}
  \end{center}
\end{figure*}

\begin{figure*}[tbp]
  \begin{center}
    \includegraphics[width=\textwidth]{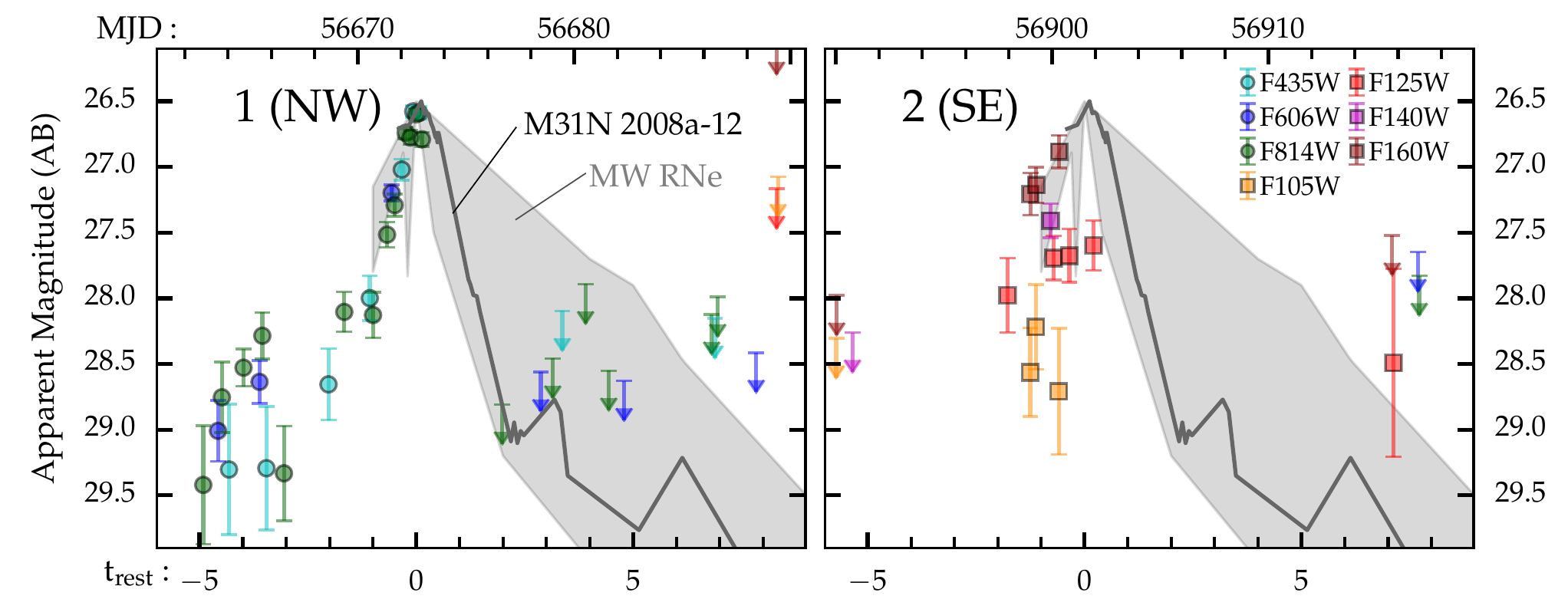}
    \caption{\protect\input{recurrent_nova_lightcurve_comparison_caption.tex}}
  \end{center}
\end{figure*}

\begin{figure*}[tbp]
  \begin{center}
    \includegraphics[width=\textwidth]{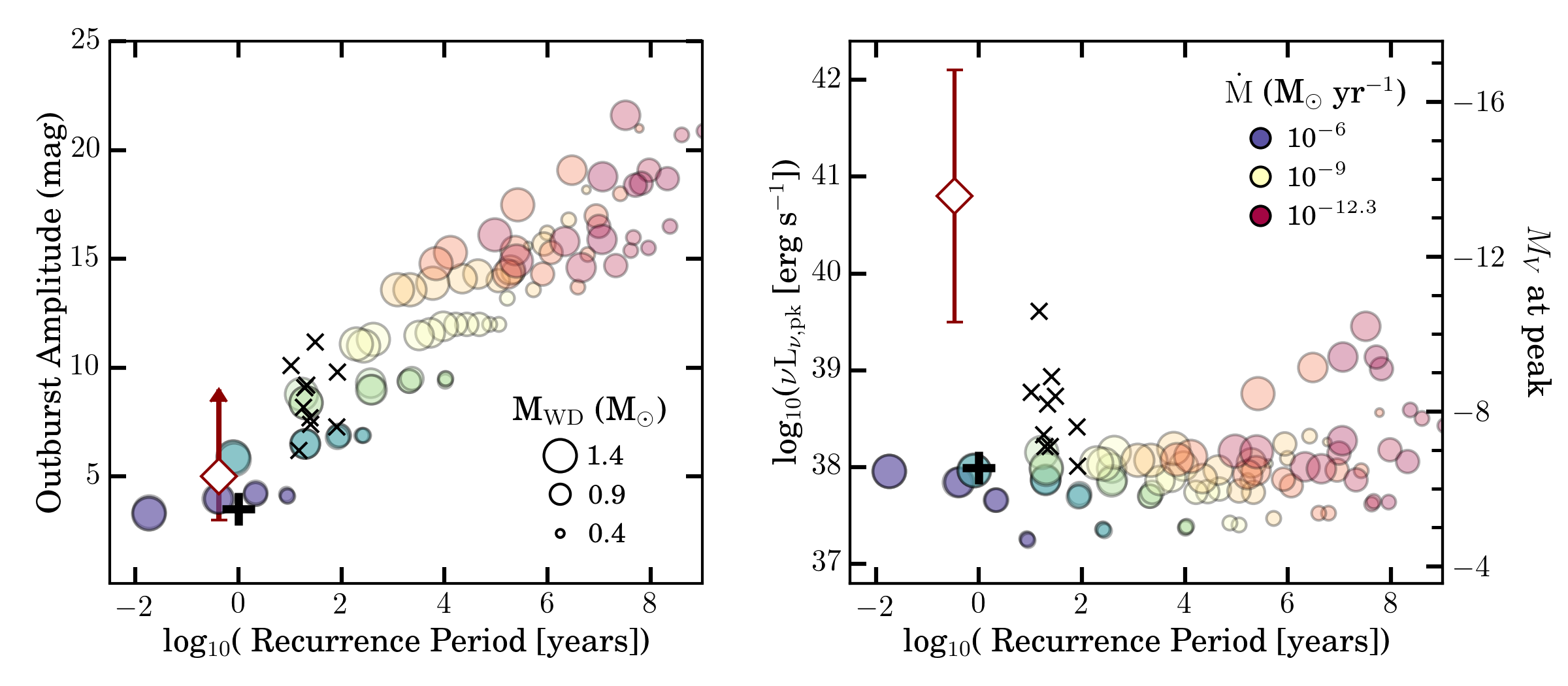}
    \caption{\protect\input{recurrent_nova_recurrence_comparison_caption.tex}}
  \end{center}
\end{figure*}

\begin{figure*}[tbp]
  \begin{center}
    \includegraphics[width=\textwidth]{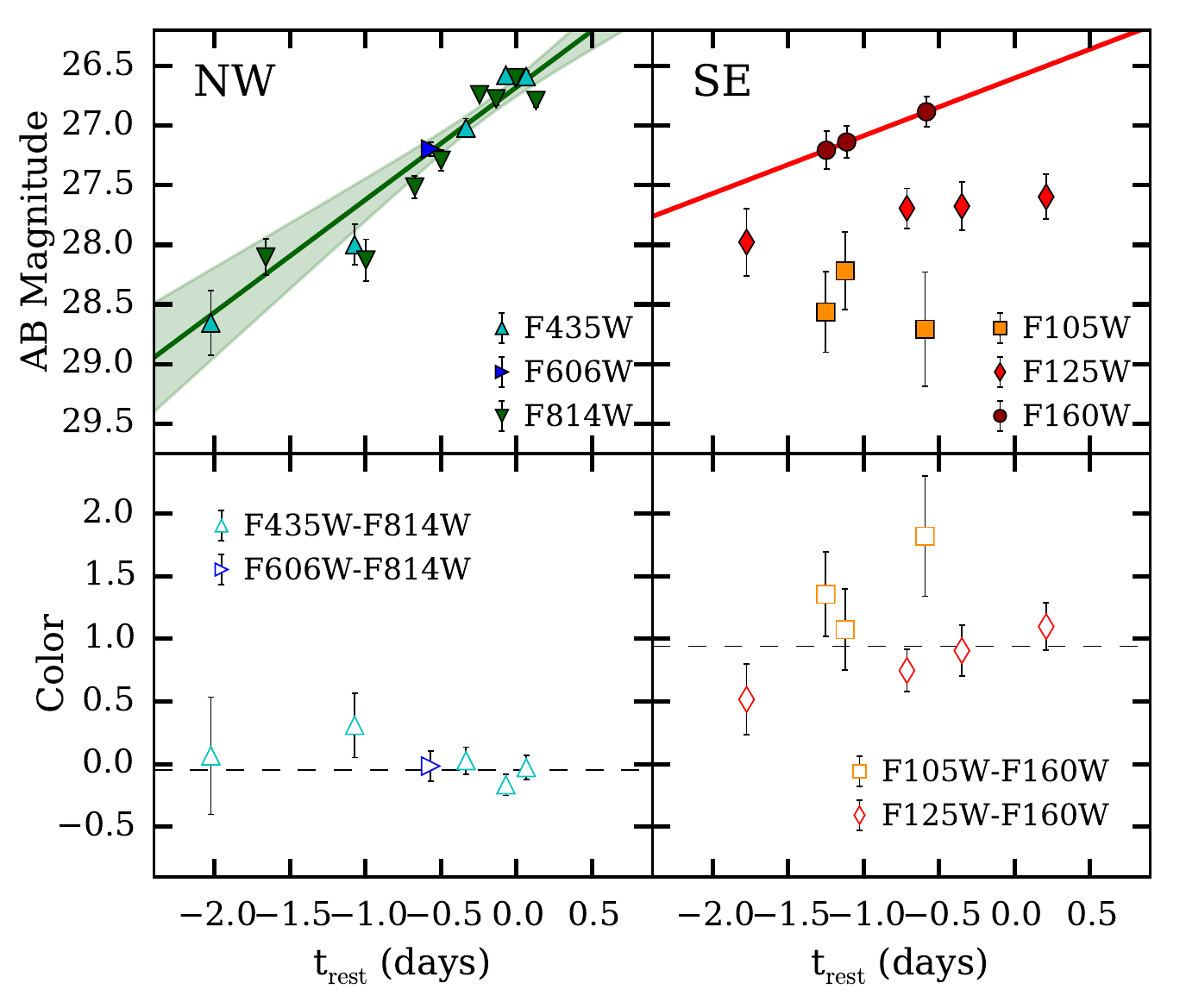}
    \caption{\protect\input{spock_colorcurves_observerframe_caption.tex}}
  \end{center}
\end{figure*}

\begin{figure*}[tbp]
  \begin{center}
    \includegraphics[width=0.45\textwidth]{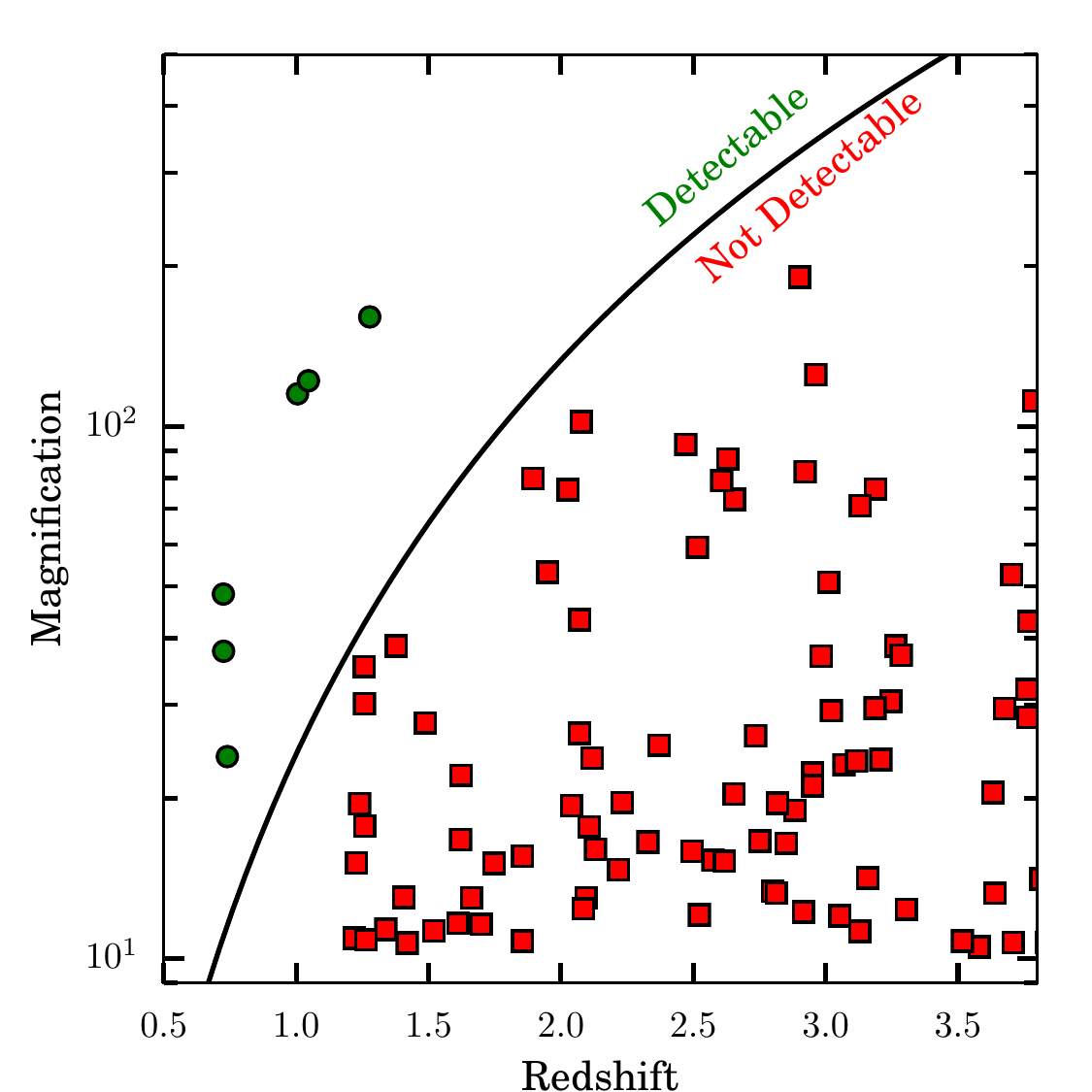}
    \caption{\protect\input{hff_lensed_galaxies_caption.tex}}
  \end{center}
\end{figure*}

\begin{deluxetable}{lccc}
  \tablewidth{0.7\linewidth}
  \tablecolumns{6}
  \tablecaption{Properties of the local stellar population in the \spock host galaxy, from SED fitting.}
  \tablehead{ {Host image:} & \colhead{11.1} & \colhead{11.2} & \colhead{11.3}\\
{Location:}   & \colhead{\spocktwo} & \colhead{\spockone} & \colhead{center}}
\startdata
$(U-V)_{\rm rest}$            & 0.69$^{+0.2}_{-0.05}$  & 0.52$^{+0.15}_{-0.10}$      & 0.39$\pm$0.05  \\
$\log[\Sigma (M_*/\Msun)]$  & 7.14 $\pm$ 0.15   & 7.14 $\pm$ 0.15     & 7.04 $\pm$ 0.10   \\
Age (Gyr)                   & 0.292$\pm$0.5 &   0.290$\pm$0.5 &  0.292$\pm$0.5  
\enddata
\label{tab:HostProperties}
\end{deluxetable}

\begin{deluxetable}{cccc}
  \tablewidth{0.7\textwidth}
  \tablecolumns{12}
  \tablecaption{Photometry of the \spockone event.\label{tab:spockonephot}}
  \tablehead{ \colhead{Date} & \colhead{Filter} & \colhead{Flux Density} & \colhead{AB Mag}\\
  \colhead{(MJD)} & \colhead{} & \colhead{(10$^{30}$ erg\,s$^{-1}$\,cm$^{-2}$\,Hz$^{-1}$)} & \colhead{} }
  \startdata
  56144.86 & F105W  &  0.063$\pm$0.119 & $<$29.39\\
56184.90 & F105W  & -0.013$\pm$0.178 & $<$27.08\\
56689.43 & F105W  & -0.111$\pm$0.179 & $<$27.08\\
56869.98 & F105W  & -0.001$\pm$0.075 & $<$28.01\\
56870.98 & F105W  &  0.068$\pm$0.048 & $<$29.32\\
56877.68 & F105W  & -0.084$\pm$0.111 & $<$27.59\\
56877.94 & F105W  &  0.005$\pm$0.051 & $<$32.09\\
56879.67 & F105W  &  0.008$\pm$0.048 & $<$31.71\\
56880.60 & F105W  & -0.003$\pm$0.057 & $<$28.32\\
56880.86 & F105W  & -0.002$\pm$0.067 & $<$28.14\\
56881.92 & F105W  & -0.065$\pm$0.100 & $<$27.71\\
56890.02 & F105W  & -0.063$\pm$0.103 & $<$27.68\\
56898.99 & F105W  & -0.005$\pm$0.079 & $<$27.96\\
56899.25 & F105W  &  0.001$\pm$0.083 & $<$34.33\\
56900.31 & F105W  &  0.014$\pm$0.067 & $<$31.02\\
56984.58 & F105W  & -0.152$\pm$0.434 & $<$26.11\\
56991.68 & F105W  & -0.242$\pm$0.396 & $<$26.21\\
57035.72 & F105W  &  0.172$\pm$0.202 & $<$28.31\\
57040.57 & F105W  &  0.460$\pm$0.211 & 27.24$\pm$0.50\\
56132.22 & F110W  &  0.066$\pm$0.046 & $<$29.35\\
56170.80 & F110W  & -0.179$\pm$0.251 & $<$26.71\\
56159.62 & F125W  & -0.124$\pm$0.208 & $<$26.91\\
56197.80 & F125W  &  0.228$\pm$0.143 & $<$28.00\\
56689.36 & F125W  &  0.028$\pm$0.165 & $<$30.27\\
56871.26 & F125W  & -0.037$\pm$0.069 & $<$28.11\\
56877.09 & F125W  &  0.010$\pm$0.048 & $<$31.43\\
56897.94 & F125W  & -0.008$\pm$0.102 & $<$27.69\\
56900.07 & F125W  & -0.063$\pm$0.116 & $<$27.55\\
56900.80 & F125W  &  0.089$\pm$0.060 & $<$29.02\\
56901.92 & F125W  &  0.061$\pm$0.056 & $<$29.43\\
56915.79 & F125W  &  0.024$\pm$0.125 & $<$30.43\\
56922.36 & F125W  &  0.393$\pm$0.360 & $<$27.41\\
56928.07 & F125W  &  0.095$\pm$0.166 & $<$28.96\\
56159.63 & F140W  & -0.144$\pm$0.256 & $<$26.69\\
56184.88 & F140W  & -0.103$\pm$0.186 & $<$27.03\\
56875.10 & F140W  & -0.026$\pm$0.067 & $<$28.14\\
56875.97 & F140W  &  0.008$\pm$0.042 & $<$31.70\\
56889.05 & F140W  & -0.019$\pm$0.081 & $<$27.94\\
56890.77 & F140W  & -0.181$\pm$0.158 & $<$27.21\\
56899.93 & F140W  & -0.126$\pm$0.148 & $<$27.28\\
56984.72 & F140W  & -0.105$\pm$0.266 & $<$26.65\\
56991.55 & F140W  &  0.202$\pm$0.160 & $<$28.14\\
57035.52 & F140W  & -0.272$\pm$0.393 & $<$26.22\\
57040.83 & F140W  &  0.004$\pm$0.239 & $<$32.32\\
56132.24 & F160W  & -0.280$\pm$0.401 & $<$26.20\\
56144.87 & F160W  &  0.001$\pm$0.267 & $<$34.03\\
56170.79 & F160W  &  0.104$\pm$0.185 & $<$28.85\\
56197.79 & F160W  & -0.239$\pm$0.332 & $<$26.40\\
56689.36 & F160W  & -0.400$\pm$0.485 & $<$25.99\\
56870.00 & F160W  &  0.083$\pm$0.067 & $<$29.10\\
56870.99 & F160W  &  0.257$\pm$0.108 & 27.88$\pm$0.46\\
56877.70 & F160W  & -0.048$\pm$0.128 & $<$27.44\\
56877.96 & F160W  & -0.008$\pm$0.101 & $<$27.70\\
56879.69 & F160W  & -0.009$\pm$0.080 & $<$27.95\\
56880.61 & F160W  &  0.035$\pm$0.069 & $<$30.05\\
56880.88 & F160W  & -0.185$\pm$0.239 & $<$26.76\\
56881.94 & F160W  & -0.070$\pm$0.141 & $<$27.34\\
56890.04 & F160W  & -0.111$\pm$0.152 & $<$27.26\\
56899.00 & F160W  & -0.064$\pm$0.141 & $<$27.33\\
56899.27 & F160W  & -0.147$\pm$0.176 & $<$27.09\\
56900.33 & F160W  & -0.014$\pm$0.122 & $<$27.50\\
56915.73 & F160W  & -0.021$\pm$0.208 & $<$26.91\\
56922.41 & F160W  & -0.120$\pm$0.313 & $<$26.47\\
56928.13 & F160W  & -0.186$\pm$0.352 & $<$26.34\\
56159.54 & F435W  &  0.012$\pm$0.016 & $<$31.23\\
56184.77 & F435W  & -0.001$\pm$0.010 & $<$30.23\\
56664.04 & F435W  &  0.069$\pm$0.031 & 29.30$\pm$0.50\\
56665.77 & F435W  &  0.070$\pm$0.030 & 29.29$\pm$0.47\\
56668.64 & F435W  &  0.125$\pm$0.031 & 28.66$\pm$0.27\\
56670.55 & F435W  &  0.229$\pm$0.036 & 28.00$\pm$0.17\\
56672.03 & F435W  &  0.565$\pm$0.041 & 27.02$\pm$0.08\\
56672.56 & F435W  &  0.849$\pm$0.035 & 26.58$\pm$0.05\\
56672.83 & F435W  &  0.840$\pm$0.038 & 26.59$\pm$0.05\\
56679.45 & F435W  & -0.065$\pm$0.070 & $<$28.10\\
56686.51 & F435W  & -0.056$\pm$0.066 & $<$28.15\\
56696.27 & F435W  & -0.030$\pm$0.065 & $<$28.18\\
56144.78 & F606W  & -0.021$\pm$0.017 & $<$30.59\\
56170.73 & F606W  & -0.117$\pm$0.110 & $<$29.89\\
56663.53 & F606W  &  0.090$\pm$0.019 & 29.01$\pm$0.23\\
56665.45 & F606W  &  0.127$\pm$0.019 & 28.64$\pm$0.16\\
56671.56 & F606W  &  0.480$\pm$0.027 & 27.20$\pm$0.06\\
56678.46 & F606W  & -0.059$\pm$0.046 & $<$28.56\\
56682.31 & F606W  & -0.060$\pm$0.043 & $<$28.63\\
56688.42 & F606W  & -0.067$\pm$0.052 & $<$28.41\\
56916.92 & F606W  & -0.020$\pm$0.064 & $<$28.20\\
56144.80 & F814W  &  0.314$\pm$0.317 & $<$27.66\\
56157.83 & F814W  &  0.028$\pm$0.041 & $<$30.29\\
56170.72 & F814W  & -0.108$\pm$0.109 & $<$27.61\\
56184.75 & F814W  &  0.004$\pm$0.007 & $<$32.47\\
56662.85 & F814W  &  0.062$\pm$0.026 & 29.42$\pm$0.45\\
56663.71 & F814W  &  0.114$\pm$0.028 & 28.75$\pm$0.27\\
56664.71 & F814W  &  0.141$\pm$0.018 & 28.53$\pm$0.14\\
56665.58 & F814W  &  0.176$\pm$0.029 & 28.28$\pm$0.18\\
56666.58 & F814W  &  0.067$\pm$0.022 & 29.33$\pm$0.36\\
56669.37 & F814W  &  0.208$\pm$0.029 & 28.10$\pm$0.15\\
56670.70 & F814W  &  0.204$\pm$0.033 & 28.13$\pm$0.17\\
56671.35 & F814W  &  0.358$\pm$0.032 & 27.52$\pm$0.10\\
56671.70 & F814W  &  0.440$\pm$0.034 & 27.29$\pm$0.09\\
56672.21 & F814W  &  0.731$\pm$0.031 & 26.74$\pm$0.05\\
56672.43 & F814W  &  0.707$\pm$0.036 & 26.78$\pm$0.06\\
56672.69 & F814W  &  0.832$\pm$0.035 & 26.60$\pm$0.05\\
56672.96 & F814W  &  0.698$\pm$0.035 & 26.79$\pm$0.06\\
56676.68 & F814W  &  0.019$\pm$0.036 & $<$30.68\\
56679.01 & F814W  & -0.015$\pm$0.050 & $<$28.46\\
56680.53 & F814W  & -0.109$\pm$0.084 & $<$27.89\\
56681.60 & F814W  & -0.021$\pm$0.046 & $<$28.55\\
56686.36 & F814W  & -0.053$\pm$0.068 & $<$28.12\\
56686.64 & F814W  & -0.092$\pm$0.077 & $<$27.99\\
56691.36 & F814W  & -0.074$\pm$0.074 & $<$28.03\\
56697.73 & F814W  & -0.040$\pm$0.051 & $<$28.43\\
56916.98 & F814W  &  0.007$\pm$0.093 & $<$31.86\\
\enddata
    \end{deluxetable}

\begin{deluxetable}{cccc}
  \tablewidth{0.7\textwidth}
  \tablecolumns{12}
  \tablecaption{Photometry of the \spocktwo event.\label{tab:spocktwophot}}
  \tablehead{ \colhead{Date} & \colhead{Filter} & \colhead{Flux Density} & \colhead{AB Mag}\\
  \colhead{(MJD)} & \colhead{} & \colhead{(10$^{30}$ erg\,s$^{-1}$\,cm$^{-2}$\,Hz$^{-1}$)} & \colhead{} }
  \startdata
  56144.86 & F105W  & -0.127$\pm$0.206 & $<$26.92\\
56184.90 & F105W  &  0.120$\pm$0.169 & $<$28.70\\
56689.43 & F105W  & -0.100$\pm$0.170 & $<$27.13\\
56869.98 & F105W  &  0.041$\pm$0.054 & $<$29.88\\
56870.98 & F105W  &  0.009$\pm$0.059 & $<$31.56\\
56877.68 & F105W  & -0.031$\pm$0.087 & $<$27.86\\
56877.94 & F105W  &  0.097$\pm$0.037 & 28.94$\pm$0.42\\
56879.67 & F105W  & -0.078$\pm$0.101 & $<$27.70\\
56880.60 & F105W  & -0.022$\pm$0.066 & $<$28.15\\
56880.86 & F105W  &  0.074$\pm$0.041 & $<$29.23\\
56881.92 & F105W  &  0.034$\pm$0.058 & $<$30.06\\
56890.02 & F105W  &  0.046$\pm$0.058 & $<$29.73\\
56898.99 & F105W  &  0.136$\pm$0.042 & 28.56$\pm$0.34\\
56899.25 & F105W  &  0.187$\pm$0.056 & 28.22$\pm$0.32\\
56900.31 & F105W  &  0.119$\pm$0.053 & 28.71$\pm$0.48\\
56984.58 & F105W  & -0.151$\pm$0.419 & $<$26.15\\
56991.68 & F105W  & -0.404$\pm$0.498 & $<$25.96\\
57035.72 & F105W  & -0.265$\pm$0.422 & $<$26.14\\
57040.57 & F105W  &  0.499$\pm$0.248 & 27.15$\pm$0.54\\
56132.22 & F110W  &  0.050$\pm$0.045 & $<$29.64\\
56170.80 & F110W  & -0.230$\pm$0.307 & $<$26.49\\
56159.62 & F125W  & -0.329$\pm$0.387 & $<$26.24\\
56197.80 & F125W  & -0.226$\pm$0.310 & $<$26.48\\
56689.36 & F125W  & -0.076$\pm$0.190 & $<$27.01\\
56871.26 & F125W  &  0.049$\pm$0.040 & $<$29.68\\
56877.09 & F125W  &  0.009$\pm$0.048 & $<$31.51\\
56897.94 & F125W  &  0.234$\pm$0.061 & 27.98$\pm$0.28\\
56900.07 & F125W  &  0.304$\pm$0.047 & 27.69$\pm$0.17\\
56900.80 & F125W  &  0.309$\pm$0.058 & 27.68$\pm$0.20\\
56901.92 & F125W  &  0.332$\pm$0.058 & 27.60$\pm$0.19\\
56915.79 & F125W  &  0.146$\pm$0.096 & $<$28.49\\
56922.36 & F125W  & -0.275$\pm$0.321 & $<$26.44\\
56928.07 & F125W  &  0.402$\pm$0.373 & $<$27.39\\
56159.63 & F140W  & -0.006$\pm$0.159 & $<$27.21\\
56184.88 & F140W  & -0.077$\pm$0.168 & $<$27.14\\
56875.10 & F140W  &  0.043$\pm$0.038 & $<$29.81\\
56875.97 & F140W  & -0.045$\pm$0.070 & $<$28.10\\
56889.05 & F140W  &  0.042$\pm$0.068 & $<$29.85\\
56890.77 & F140W  &  0.055$\pm$0.060 & $<$29.55\\
56899.93 & F140W  &  0.395$\pm$0.047 & 27.41$\pm$0.13\\
56984.72 & F140W  & -0.226$\pm$0.320 & $<$26.44\\
56991.55 & F140W  &  0.003$\pm$0.184 & $<$32.56\\
57035.52 & F140W  & -0.068$\pm$0.255 & $<$26.69\\
57040.83 & F140W  &  0.142$\pm$0.174 & $<$28.52\\
56132.24 & F160W  & -0.196$\pm$0.355 & $<$26.33\\
56144.87 & F160W  & -0.592$\pm$0.716 & $<$25.57\\
56170.79 & F160W  & -0.375$\pm$0.461 & $<$26.05\\
56197.79 & F160W  & -0.766$\pm$0.725 & $<$25.56\\
56689.36 & F160W  & -0.282$\pm$0.405 & $<$26.19\\
56870.00 & F160W  &  0.028$\pm$0.069 & $<$30.30\\
56870.99 & F160W  &  0.162$\pm$0.059 & 28.38$\pm$0.39\\
56877.70 & F160W  & -0.047$\pm$0.131 & $<$27.42\\
56877.96 & F160W  & -0.002$\pm$0.095 & $<$27.76\\
56879.69 & F160W  &  0.004$\pm$0.086 & $<$32.34\\
56880.61 & F160W  & -0.043$\pm$0.113 & $<$27.57\\
56880.88 & F160W  &  0.126$\pm$0.065 & $<$28.65\\
56881.94 & F160W  & -0.026$\pm$0.125 & $<$27.46\\
56890.04 & F160W  &  0.035$\pm$0.078 & $<$30.05\\
56899.00 & F160W  &  0.476$\pm$0.070 & 27.21$\pm$0.16\\
56899.27 & F160W  &  0.507$\pm$0.063 & 27.14$\pm$0.13\\
56900.33 & F160W  &  0.640$\pm$0.074 & 26.88$\pm$0.12\\
56915.73 & F160W  &  0.053$\pm$0.119 & $<$29.59\\
56922.41 & F160W  &  0.037$\pm$0.235 & $<$29.99\\
56928.13 & F160W  & -0.281$\pm$0.466 & $<$26.04\\
56159.54 & F435W  &  0.002$\pm$0.007 & $<$33.06\\
56184.77 & F435W  & -0.009$\pm$0.013 & $<$29.93\\
56664.04 & F435W  &  0.033$\pm$0.034 & $<$30.11\\
56665.77 & F435W  & -0.019$\pm$0.039 & $<$28.73\\
56668.64 & F435W  &  0.044$\pm$0.035 & $<$29.79\\
56670.55 & F435W  &  0.047$\pm$0.043 & $<$29.71\\
56672.03 & F435W  &  0.051$\pm$0.039 & $<$29.63\\
56672.56 & F435W  & -0.046$\pm$0.065 & $<$28.18\\
56672.83 & F435W  & -0.064$\pm$0.075 & $<$28.01\\
56679.45 & F435W  &  0.018$\pm$0.035 & $<$30.74\\
56686.51 & F435W  & -0.026$\pm$0.052 & $<$28.41\\
56696.27 & F435W  &  0.009$\pm$0.049 & $<$31.48\\
56144.78 & F606W  &  0.234$\pm$0.176 & $<$27.98\\
56170.73 & F606W  &  0.043$\pm$0.061 & $<$29.81\\
56663.53 & F606W  &  0.004$\pm$0.027 & $<$32.27\\
56665.45 & F606W  &  0.016$\pm$0.026 & $<$30.91\\
56671.56 & F606W  & -0.021$\pm$0.043 & $<$28.61\\
56678.46 & F606W  & -0.012$\pm$0.029 & $<$29.05\\
56682.31 & F606W  & -0.016$\pm$0.035 & $<$28.83\\
56688.42 & F606W  &  0.033$\pm$0.024 & $<$30.10\\
56916.92 & F606W  & -0.121$\pm$0.106 & $<$27.65\\
56144.80 & F814W  & -0.294$\pm$0.274 & $<$26.61\\
56157.83 & F814W  & -0.025$\pm$0.146 & $<$27.29\\
56170.72 & F814W  & -0.027$\pm$0.127 & $<$27.45\\
56184.75 & F814W  &  0.037$\pm$0.010 & 29.97$\pm$0.29\\
56662.85 & F814W  &  0.026$\pm$0.036 & $<$30.37\\
56663.71 & F814W  & -0.019$\pm$0.052 & $<$28.42\\
56664.71 & F814W  &  0.005$\pm$0.026 & $<$32.15\\
56665.58 & F814W  & -0.056$\pm$0.072 & $<$28.07\\
56666.58 & F814W  & -0.007$\pm$0.038 & $<$28.75\\
56669.37 & F814W  &  0.010$\pm$0.042 & $<$31.44\\
56670.70 & F814W  &  0.005$\pm$0.042 & $<$32.12\\
56671.35 & F814W  &  0.075$\pm$0.029 & 29.21$\pm$0.41\\
56671.70 & F814W  &  0.209$\pm$0.156 & $<$28.10\\
56672.21 & F814W  & -0.014$\pm$0.054 & $<$28.38\\
56672.43 & F814W  &  0.005$\pm$0.041 & $<$32.14\\
56672.69 & F814W  & -0.003$\pm$0.036 & $<$28.81\\
56672.96 & F814W  & -0.020$\pm$0.053 & $<$28.41\\
56676.68 & F814W  & -0.005$\pm$0.042 & $<$28.64\\
56679.01 & F814W  &  0.016$\pm$0.040 & $<$30.89\\
56680.53 & F814W  & -0.009$\pm$0.045 & $<$28.57\\
56681.60 & F814W  &  0.012$\pm$0.037 & $<$31.22\\
56686.36 & F814W  & -0.031$\pm$0.055 & $<$28.35\\
56686.64 & F814W  & -0.031$\pm$0.050 & $<$28.45\\
56691.36 & F814W  &  0.042$\pm$0.036 & $<$29.84\\
56697.73 & F814W  &  0.064$\pm$0.019 & 29.39$\pm$0.31\\
56916.98 & F814W  & -0.028$\pm$0.089 & $<$27.83\\
\enddata
    \end{deluxetable}

\end{supplementary}

\renewcommand\refname{References for Supplementary Information}

\end{document}